\documentclass[fleqn,usenatbib]{mnras}

\usepackage{newtxtext,newtxmath}

\usepackage[T1]{fontenc}

\DeclareRobustCommand{\VAN}[3]{#2}
\let\VANthebibliography\thebibliography
\def\thebibliography{\DeclareRobustCommand{\VAN}[3]{##3}\VANthebibliography}

\usepackage{graphicx}	
\usepackage{amsmath}	
\usepackage{bm}
\usepackage{booktabs}
\usepackage{threeparttable}
\usepackage{xcolor}
\usepackage[normalem]{ulem}
\usepackage[export]{adjustbox}

\newcommand{\email}[1]{\mbox{\href{mailto:#1}{#1}}}

\makeatletter
\newlength{\abovecaptionskip}%
\makeatother

\defcitealias{mandel20}{M20}
\defcitealias{brout20}{BS21}
\defcitealias{burns14}{B14}

\DeclareMathOperator{\exponential}{Exponential}

\DeclareMathOperator{\gammadist}{Gamma}

\title[Dust Laws and the Mass Step with BayeSN]{Constraining the SN Ia Host Galaxy Dust Law Distribution and Mass Step: Hierarchical BayeSN Analysis of Optical and Near-Infrared Light Curves}

\author[S.\ Thorp \& K.~S.\ Mandel]{Stephen Thorp$^{1}$\thanks{E-mail: \email{sjt202@cam.ac.uk}} and Kaisey S.\ Mandel$^{1,2,3}$\\
$^1$Institute of Astronomy and Kavli Institute for Cosmology, Madingley Road, Cambridge, CB3 0HA, UK\\
$^2$Statistical Laboratory, DPMMS, University of Cambridge, Wilberforce Road, Cambridge, CB3 0WB, UK\\
$^3$The Alan Turing Institute, Euston Road, London, NW1 2DB, UK\\
}

\date{Accepted XXX. Received YYY; in original form ZZZ}

\pubyear{2022}

\begin{document}
\label{firstpage}
\pagerange{\pageref{firstpage}--\pageref{lastpage}}
\maketitle

\begin{abstract}
We use the BayeSN hierarchical probabilistic SED model to analyse the optical--NIR ($BVriYJH$) light curves of 86 Type Ia supernovae (SNe Ia) from the Carnegie Supernova Project to investigate the SN~Ia host galaxy dust law distribution and correlations between SN~Ia Hubble residuals and host mass. Our Bayesian analysis simultaneously constrains the mass step and dust $R_V$ population distribution by leveraging optical--NIR colour information. We demonstrate how a simplistic analysis where individual $R_V$ values are first estimated for each SN separately, and then the sample variance of these point estimates is computed, overestimates the $R_V$ population variance $\sigma_R^2$. This bias is exacerbated when neglecting residual intrinsic colour variation beyond that due to light curve shape. Instead, Bayesian shrinkage estimates of $\sigma_R$ are more accurate, with fully hierarchical analysis of the light curves being ideal. For the 75 SNe with low-to-moderate reddening (peak apparent $B-V\leq0.3$), we estimate an $R_V$ distribution with population mean $\mu_R=2.59\pm0.14$, and standard deviation $\sigma_R=0.62\pm0.16$. Splitting this subsample at the median host galaxy mass ($10^{10.57}~\mathrm{M}_\odot$) yields consistent estimated $R_V$ distributions between low- and high-mass galaxies, with $\mu_R=2.79\pm0.18$, $\sigma_R=0.42\pm0.24$, and $\mu_R=2.35\pm0.27$, $\sigma_R=0.74\pm0.36$, respectively. When estimating distances from the full optical--NIR light curves while marginalising over various forms of the dust $R_V$ distribution, a mass step of $\gtrsim0.06$~mag persists in the Hubble residuals at the median host mass.
\end{abstract}




\section{Introduction}
One of the most well-known correlations between Type Ia supernovae (SNe Ia) and their host galaxies is the so-called ``mass step''. This is an empirical effect whereby SNe Ia  in high mass galaxies are observed to be systematically brighter, post-standardisation, than their counterparts in lower mass hosts \citep{kelly10, sullivan10}. Many past studies \citep[e.g.][]{sullivan10, childress14, kim18, rigault20, briday21} have linked this to a difference in the populations of SN Ia progenitors between more and less massive host galaxies, something that could arise from a ``prompt/delayed'' progenitor model \citep[\`a la][]{mannucci05,mannucci06,scannapieco05}. Such a host-dependent progenitor scenario could also explain observed correlations between light curve shape and host properties \citep[see e.g.][]{hamuy96, hamuy00, sullivan06, childress14, nicolas21}. Alternatively, it has recently been proposed that the mass step can be explained by a difference in host galaxy properties, based on analyses at optical wavelengths \citep{brout20, popovic21}. However, recent findings that the mass step may extend to the near-infrared (NIR) would seem to be in tension with this \citep{ponder20, uddin20, jones22}, although the strength of any NIR mass step is currently uncertain \citep{ponder20, johansson21, jones22}. The interplay between dust and host mass \citep[see e.g.][]{brout20, gonzalezgaitan21, thorp21, johansson21, popovic21, meldorf22, wiseman22}, and the distribution of dust properties in SN Ia hosts more broadly, remains controversial, and the problem of accurately estimating the effect of dust on SN Ia observations is a challenging one. In this paper, we use our \textsc{BayeSN} model \citep{thorp21,mandel20} for the optical--NIR spectral energy distributions (SEDs) of SNe Ia to place constraints on the SN Ia host galaxy dust law distribution via an analysis of optical and NIR data from the Carnegie Supernova Project. In parallel to this, we investigate the sensitivity of the mass step to a variety of assumptions about the distribution of host galaxy dust laws. We also present an argument for the importance of high quality optical and NIR data, and a simulation-based study demonstrating some of the challenges inherent in estimating the dust law distribution, and how these can be overcome using hierarchical Bayes. 

The SN Ia mass step \citep{kelly10, sullivan10} is an observed tendency for SNe Ia in high mass galaxies to have brighter post-standardisation magnitudes (or more negative Hubble residuals) than those in low-mass host galaxies. In optical magnitudes (or in Hubble residuals derived from optical light curves), a step of $\sim0.04$--$0.1$~mag has typically been observed \citep{kelly10, sullivan10, childress13, betoule14, roman18, scolnic18, jones18, jones19, smith20, kelsey21}, located at host masses between $\sim10^{10}~\mathrm{M}_\odot$ \citep[e.g.][]{sullivan10} and $10^{10.8}~\mathrm{M}_\odot$ \citep[e.g.][]{kelly10}. Magnitude steps versus other host galaxy properties -- particularly (local) specific star formation rate \citep{lampeitl10,rigault13, rigault15, rigault20, jones18, kim18, briday21} -- have also been observed. Although empirically well studied in the optical, the possibility of a NIR mass step has only been investigated in a small number of recent works \citep{ponder20,uddin20,johansson21,jones22}. 

\citet{ponder20} assembled a sample of 143 SNe Ia with NIR light curves, based on a compilation \citep{weyant14} sourced from the CfA/CfAIR Supernova Survey \citep{woodvasey08, friedman15}, Carnegie Supernova Project \citep[CSP-I;][]{contreras10, stritzinger11}, SweetSpot \citep{weyant14,weyant18}, the \citet{baronenugent12} sample,  and elsewhere \citep{jha99, hernandez00, krisciunas00, krisciunas03, krisciunas04a, krisciunas04b, krisciunas07, valentini03, phillips06, pastorello07a, pastorello07b, stanishev07, pignata08}. They find that the peak $H$-band magnitudes of their sample tend to be brighter for the supernovae in more massive host galaxies. For a step in $H$-band magnitude at a host stellar mass of $10^{10.43}~\mathrm{M}_\odot$ (preferred based on the Akaike Information Criterion), they find a magnitude difference of $0.13\pm0.04$. Omitting prominent outliers reduces this to $0.08\pm0.04$~mag, at a preferred step location of $10^{10.65}~\mathrm{M}_\odot$. For the same set of supernovae, they find that the Hubble residuals computed from stretch- and colour-corrected optical data favour a step at $10^{10.65}~\mathrm{M}_\odot$, with a size of $0.14\pm0.04$~mag (and at $10^{10.43}~\mathrm{M}_\odot$, they estimate an optical mass step of $0.10\pm0.03$~mag). 

\citet{uddin20} observe a similar effect in a sample of 113 SNe Ia taken from CSP-I \citep{krisciunas17}. Fitting their light curves using \textsc{SNooPy} \citep{burns11}, they infer a distance estimate (and thus a Hubble residual) for each supernova in each passband (of \textit{uBgVriYJH}). Then, assuming a step in Hubble residual located at their median host stellar mass ($\sim10^{10.48}~\mathrm{M}_\odot)$, they compute step sizes by taking the weighted average of the Hubble residuals either side of their step. This yields a step of $\gtrsim2\sigma$ across all passbands, including in the NIR. Their $H$-band step is of size $0.093\pm0.043$~mag, consistent with that of \citet{ponder20}. In the $J$-band, they estimate a step of $0.090\pm0.046$~mag.

In their NIR mass step analysis, \citet{johansson21} combine a sample of $\sim200$ literature SNe Ia (mainly CfA/CfAIR and CSP-I, plus others from \citealp{baronenugent12,stanishev18,amanullah15}) with $\sim40$ more observed as part of their own survey using the intermediate Palomar Transient Factory \citep[iPTF;][]{rau09} and Reionization and Transients InfraRed Camera \citep[RATIR;][]{butler12}. They split their sample at $10^{10}~\mathrm{M}_\odot$, and used \textsc{SNooPy} \citep{burns11} to fit the optical and NIR light curves. With a fixed host galaxy dust law $R_V=2.0$ (close to the weighted sample mean of $\mu_R=1.9$ that they estimate from individual \textsc{SNooPy} fits to all $0.06<E(B-V)<0.5$ SNe), they estimate NIR mass steps of $0.021\pm0.033$~mag in the $J$-band, and $-0.020\pm0.036$~mag in the $H$-band -- both consistent with zero (although also within 1.2 and $2.0\sigma$, respectively, of the $J$- and $H$-band steps estimated by \citealp{uddin20}). This is contrasted with a step of $0.070\pm0.030$~mag estimated from the optical ($BgVri$), and a step very similar in size to this in the $Y$-band (see their fig.\ 13). When using an individually-fitted host galaxy $R_V$ for every supernova, they claim a mass step consistent with zero across the full optical--NIR ($BgVriYJH$) range \citep[see][fig.\ 13]{johansson21}.

Most recently, \citet{jones22} combined a sample of 42 low-redshift ($z<0.1$) SNe Ia from CSP-I, with 37 higher-redshift ($0.2<z<0.7$) SNe Ia with rest frame NIR observations obtained for the RAISIN programme using the \textit{Hubble Space Telescope} (\textit{HST}). Using the NIR light curves of their combined sample of 79 SNe, they estimate a mass step of $0.072\pm0.041$~mag at $10^{10}~\mathrm{M}_\odot$. Placing the step instead at a host mass of $10^{10.44}~\mathrm{M}_\odot$ \citep[the best-fitting step location reported by][]{ponder20}, they estimate a slightly smaller step of $0.057\pm0.035$~mag. They find similar sized steps when using the high-$z$ \textit{HST} data alone, but these are more uncertain. From optical data, they find a Hubble residual step of $0.11\pm0.03$~mag using \textsc{SNooPy} \citep{burns11}, and $0.08\pm0.03$~mag using SALT2 \citep{guy07,guy10,betoule14} or SALT3 \citep{kenworthy21}.

The possible extension of the mass step into the NIR has important implications for the underlying physical cause of SN--host correlations. It has been suggested by \citet{brout20} and \citet{popovic21} that differing distributions of the host galaxy dust law $R_V$ in galaxies greater or less massive than $10^{10}~\mathrm{M}_\odot$ can explain the conventional optical mass step. However, the detection of a significant mass step in the NIR, where sensitivity to dust extinction (and particularly to the value of $R_V$) is reduced, would seem at odds with such an explanation \citep{ponder20, uddin20}\footnote{Although, it is worth acknowledging that the non-detection of a NIR mass step does not necessarily imply that dust \textit{does} explain the mass step. It could simply be that intrinsic differences between SN Ia populations in in low- and high-mass hosts are less prominent in the NIR. Such a result might not be surprising, given the generally lower intrinsic variation seen amongst normal SNe Ia observed in the NIR \citep[see e.g.][]{avelino19}.}. If the variation of dust properties with host galaxy mass cannot adequately explain the mass step, it would suggest that host mass traces (perhaps imperfectly) something about the intrinsic properties of SNe Ia. Whatever the case, any physical explanation for the mass step must be able to explain the fact that such steps have been observed in the literature following a variety of different SN Ia standardisation approaches, and when splitting at a range of different host masses between $\sim10^{10}$ and $10^{10.8}~\mathrm{M}_\odot$ \citep[e.g.][]{kelly10, sullivan10, childress13, uddin20, smith20, kelsey21, ponder20}.

The interpretation of the mass as a tracer of intrinsic SN Ia properties has recently been supported by \citet{briday21}, in their analysis of 110 low redshift SNe Ia observed by the Nearby Supernova Factory \citep[SNfactory;][]{aldering02,aldering20,rigault20}. They found that the data were well modelled as coming from two underlying populations, best traced by local specific star formation rate (lsSFR) or total host galaxy stellar mass, and differing in post-standardization magnitude by $\gtrsim0.12$~mag. Additionally, they combine literature reports of magnitude steps versus different host properties (lsSFR; \citealp{rigault20}; $\text{UV}-\text{optical}$ colour; \citealp{roman18, kelsey21}; global mass; \citealp{roman18, smith20}; local mass; \citealp{jones18, kelsey21}; morphology; \citealp{pruzhinskaya20}) with their own equivalent results, concluding that the different step sizes follow naturally from their estimates of how well different host properties delineate between underlying SN Ia populations. Given that they find lsSFR to be the strongest environmental tracer of a magnitude step, they surmise that progenitor age is likely the fundamental property separating the two SN Ia populations that give rise to the mass step. This agrees with previous magnitude step studies \citep[e.g.][]{rigault13, rigault20, roman18, kim18}, and would align well with a ``prompt/delayed'' progenitor model \citep[see e.g.][]{mannucci05,mannucci06,scannapieco05}.

A point of contention in recent studies of correlations between SNe Ia and their hosts has been the impact of host galaxy dust -- particularly the distribution of the dust law $R_V$ parameter. Along sight lines within the Milky Way, $R_V$ has been found to follow a nearly Gaussian distribution \citep[fig.\ 15]{schlafly16} with small dispersion ($\sigma_R\approx0.2$) and a mean of $\mu_R\approx3.3$. Along sight lines in SN Ia host galaxies, a definitive constraint on the $R_V$ distribution has proven elusive. In their study of the mass step, \citet{brout20} found that SN Ia survey simulations with wide ($\sigma_R=1.3\pm0.2$) Gaussian population distributions of $R_V$ (with means of $2.75\pm0.35$ and $1.50\pm0.25$, respectively, in low- and high-mass hosts) provided the best match to the SALT2 \citep{guy07,guy10} fits of the optical light curves of a sample of 1445 real SNe Ia (from CSP-I; the CfA; \citealp{hicken09,hicken12}; Foundation; \citealp{foley18,jones19}; the Pan-STARRS-1 Medium Deep Survey \citealp{rest14,scolnic18}; Supernova Legacy Survey; \citealp{astier06,betoule14}; SDSS-II; \citealp{frieman08,sako11,sako18}; and the Dark Energy Survey; \citealp{des16,brout19}). A limitation of this approach lies in the use of \textsc{SALT2}'s $c$ parameter (roughly equivalent to apparent $B-V$ colour at peak) as a single proxy for all the colour information in the full SN Ia light curves. When the data are reduced to such a summary, considerable care must be taken to avoid a confounding of the intrinsic colour distribution and the effect of dust \citep[see][]{mandel17}. A related limitation is that SALT2 employs a single fixed colour law that translates the apparent colour parameter into an effect on the SED, and so it cannot properly accommodate the impact of variable dust laws on the SED. 

The physical effect of dust in the narrow optical $B-V$ wavelength range also should produce effects consistent with dust over a wider span of wavelengths. Hence, for reliable inference of $R_V$, it is preferable to model the the effect of dust on the full light curves and their underlying SEDs directly, ideally in both the optical and NIR, thereby leveraging a wider range of colour information across the full time-varying SED in order to break degeneracies \citep[see e.g.][]{krisciunas07,mandel11,mandel20}. This is the approach we take in this paper.

In their recent study, \citet{johansson21} also investigated the trade-off between the mass step and the distribution of $R_V$ in SN Ia hosts. Their analysis used the \textsc{SNooPy} light curve model \citep{burns11} to fit the full range of optical and NIR data. However, they estimate the population distribution of $R_V$ from a collection of independent fits to their sample of SNe Ia, taking the best-fitting values of $R_V$ for each individual and constructing a distribution from these. A statistical pitfall inherent in using a collection of individual $R_V$ point estimates is that these will likely be over-dispersed, due to the individual estimation errors, compared to the underlying population distribution, meaning the width of the population distribution can easily be overestimated by taking the sample standard deviation of these. This can be remedied by adopting a hierarchical Bayesian approach (see e.g. \citealp{mandel11}, \citealp{burns14}, \citealp{thorp21}, for examples in this context; or \citealp{loredo10,loredo19}, for a pedagogical discussion). However, difficulty can still arise if the diversity of SN Ia intrinsic colours is not included within the model used to estimate $R_V$ (e.g.\ the \textsc{SNooPy} \texttt{color\_model} used by \citealp{johansson21} does not account for any residual intrinsic colour variation beyond the stretch--colour relationship inferred in \citealp{burns14}). Not allowing for the intrinsic colour scatter of SNe Ia is known to bias $R_V$ estimates \citep[see][]{mandel17}, and could easily lead one to overestimate the amount of apparent colour dispersion that should be attributed to a range of $R_V$ values \citep[see discussion in][\S4.3.2]{thorp21}. We will explore these challenges in more depth later in this paper (in \S \ref{sec:why_hb}), and will demonstrate how they can be alleviated.

A hierarchical Bayesian approach to the modelling and inference of the dust law $R_V$ distribution in SN Ia host galaxies was first developed, and applied to a sample of optical and NIR light curves, by \citet{mandel11}. The \textsc{BayeSN} light curve modelling framework of \citet{mandel09,mandel11} was recently generalised to the modelling of SN Ia SEDs by \citet{mandel20}, and applied by \citet{thorp21} to constrain the distribution of dust laws in the host galaxies of 157 SNe Ia from the Foundation Supernova Survey \citep{foley18,jones19}. In that analysis, \citet{thorp21} allowed for different $R_V$ population distributions in low- and high-mass host galaxies, including this within their hierarchical model alongside a conventional host galaxy mass step. Unlike \citet{brout20}, \citet{thorp21} concluded that the distribution of dust laws was not significantly different in low- and high-mass hosts, and that a scenario where a large offset in mean $R_V$ replaced the conventional mass step was not favoured. Their results also favoured an overall small dispersion in $R_V$ across their sample (with population standard deviation $\lesssim0.6$ with 95 per cent posterior probability). Their analysis was able to leverage all available phototometry from the $g$- through to the $z$-band ($3500\lesssim\lambda\lesssim9500$~\AA), within their hierarchical Bayesian framework, to inform the dust constraints. However, the inclusion of data further into the NIR, and the consideration of SNe Ia with redder apparent colours ($B-V>0.3$), has the potential to yield much stronger constraints and new insights.

In this work, we build on the advances made by \citet{thorp21} and present an analysis of 86 SNe Ia from CSP-I \citep{krisciunas17}. We investigate the strength of the host galaxy mass step in the Hubble residuals for this sample. Our analysis is carried out using \textsc{BayeSN} fits to the full optical+NIR light curves, and to the NIR light curves alone. We focus particularly on whether the treatment of host galaxy dust laws impacts the mass step in Hubble residuals estimated from joint fits to the optical and NIR. We are able to draw conclusions about both the mass step, and the host galaxy dust law $R_V$ population distribution. Using the CSP sample enables us to leverage a wider wavelength range ($BVriYJH$ passbands; $\approx$3500--18000~\AA) than was possible with the Foundation data ($griz$ passbands; $\approx$3500--9500~\AA) used in \citet{thorp21}, enabling stronger constraining power for the dust law $R_V$ values of individual SNe. Additionally, we are able to include more highly reddened supernovae, allowing us to explore the possibility that $R_V$ evolves with reddening/extinction \citep[also studied by][]{mandel11,burns14}. We also verify that our results are robust to the assumption of some non-Gaussian forms for the $R_V$ population distribution.

In Section \ref{sec:photodata}, we describe the sample of 86 CSP SNe Ia used in this work. In Section \ref{sec:fitting}, we describe the \textsc{BayeSN} SN Ia SED model (\S\ref{sec:bayesn}), the different photometric distance fitting procedures we adopt (\S\ref{sec:fittingconfigs}), the posterior distributions that we sample in each case (\S\ref{sec:jointpost}), our estimation of Hubble residuals from the photometric distances (\S\ref{sec:resids}), and our approach to estimating the size of the mass step in a set of Hubble residuals (\S\ref{sec:searching}). Before presenting our analysis of the CSP data, we include two sections motivating the benefit of combining optical and NIR data for constraining dust (\S\ref{sec:why_onir}), and the benefit of the hierarchical Bayesian approach (\S\ref{sec:why_hb}). In Section \ref{sec:results}, we describe our results, with Section \ref{sec:step} focusing on the Hubble residual step, Section \ref{sec:dustlaws} covering our inferences about the $R_V$ population distribution, Section \ref{sec:nirmax} presenting results on a gold-standard SN Ia subsample with NIR data near maximum, and Section \ref{sec:highreddening} discussing our $R_V$ inferences for the most highly reddened SNe Ia. We offer final discussion and conclusions in Section \ref{sec:conclusions}.

\section{Data}
\label{sec:photodata}
In this work, we use a sample of 86 spectroscopically normal SNe Ia with optical and NIR light curves from CSP-I \citep{krisciunas17}. All have self-consistent mass estimates, obtained using Z-PEG \citep{leborgne02}, from \citet{uddin20}. Our sample of 86 is a subset of the 113 supernovae used in the analysis of \citet{uddin20}. The \citet{uddin20} sample is the full CSP sample (134 SNe), minus 13 SNe deemed to be peculiar, 7 with ambiguous hosts, and 1 with a poor quality light curve \citep[see][\S2.1--2.2 for details]{uddin20}. Of the 113 SNe in the \citet{uddin20} sample, 100 have near-infrared light curves (those without are: 2005W, 2005be, 2005bg, 2005bl, 2005bo, 2005ir, 2005mc, 2006ef, 2006fw, 2006py, 2007ol, 2008bz, 2008cd). We apply a stricter cut on spectroscopic normality, bringing this down to 86. Compared to the 100 SNe with NIR light curves in \citet{uddin20}, we have eliminated 9 SNe (2004gu, 2005ke, 2006bd, 2006gt, 2006mr, 2007N, 2007ax, 2007ba, 2009F; mostly 91bg-like) identified as Ia-pec in \citet[table 3]{friedman15}, and 5 SNe identified as 91bg-like (2007al, 2008bi, 2008bt), 86G-like (2007jh), or not subtyped (2007cg) in \citet{krisciunas17}. We analyse this sample of 86 SNe Ia, and its partitions, for the remainder of this work. Table \ref{tab:sample_cuts} summarises the cuts leading to this sample. 

\begin{table}
\centering
    \begin{threeparttable}
        \caption{Sample cuts.}
        \label{tab:sample_cuts}
        \begin{tabular}{l r}\toprule
            Cut & $N_\text{SN}$ remaining \\\midrule
            Full CSP-I sample & 134\\
            \citet{uddin20} sample\tnote{a} & 113\\
            NIR data & 100\\
            Spectroscopically\ normal\tnote{b} & 86\\
            \bottomrule
        \end{tabular}
        \begin{tablenotes}
            \item [a] Ambiguous hosts, and some deemed peculiar.
            \item [b] SNe identified as peculiar (91bg-, 06gz-, or 86G-like) in \citep[tab.\ 2]{krisciunas17} or \citep[tab.\ 3]{friedman15}.
        \end{tablenotes}
    \end{threeparttable}
\end{table}

\begin{figure}
    \centering
    \includegraphics[width=\linewidth]{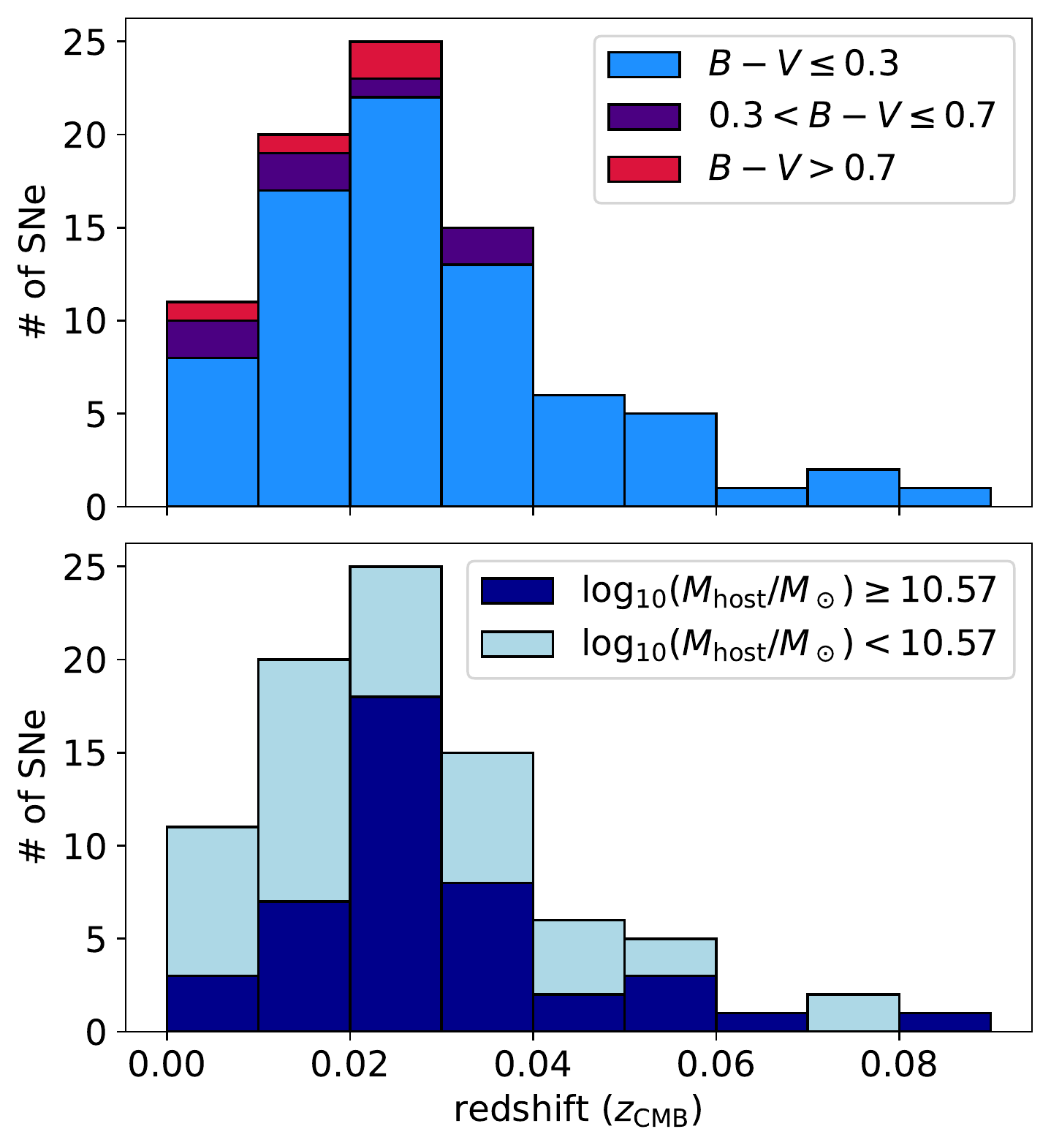}
    \caption{CMB-frame redshift distribution of the 86 SNe Ia used in this work. Bins have a width of 0.01. (top panel) Bins subdivided by apparent $B-V$ colour. (bottom panel) Bins subdivided by host galaxy stellar mass (at the sample median, $10^{10.57}~\mathrm{M}_\odot$).}
    \label{fig:redshiftdist}
\end{figure}

\begin{figure}
    \centering
    \includegraphics[width=\linewidth]{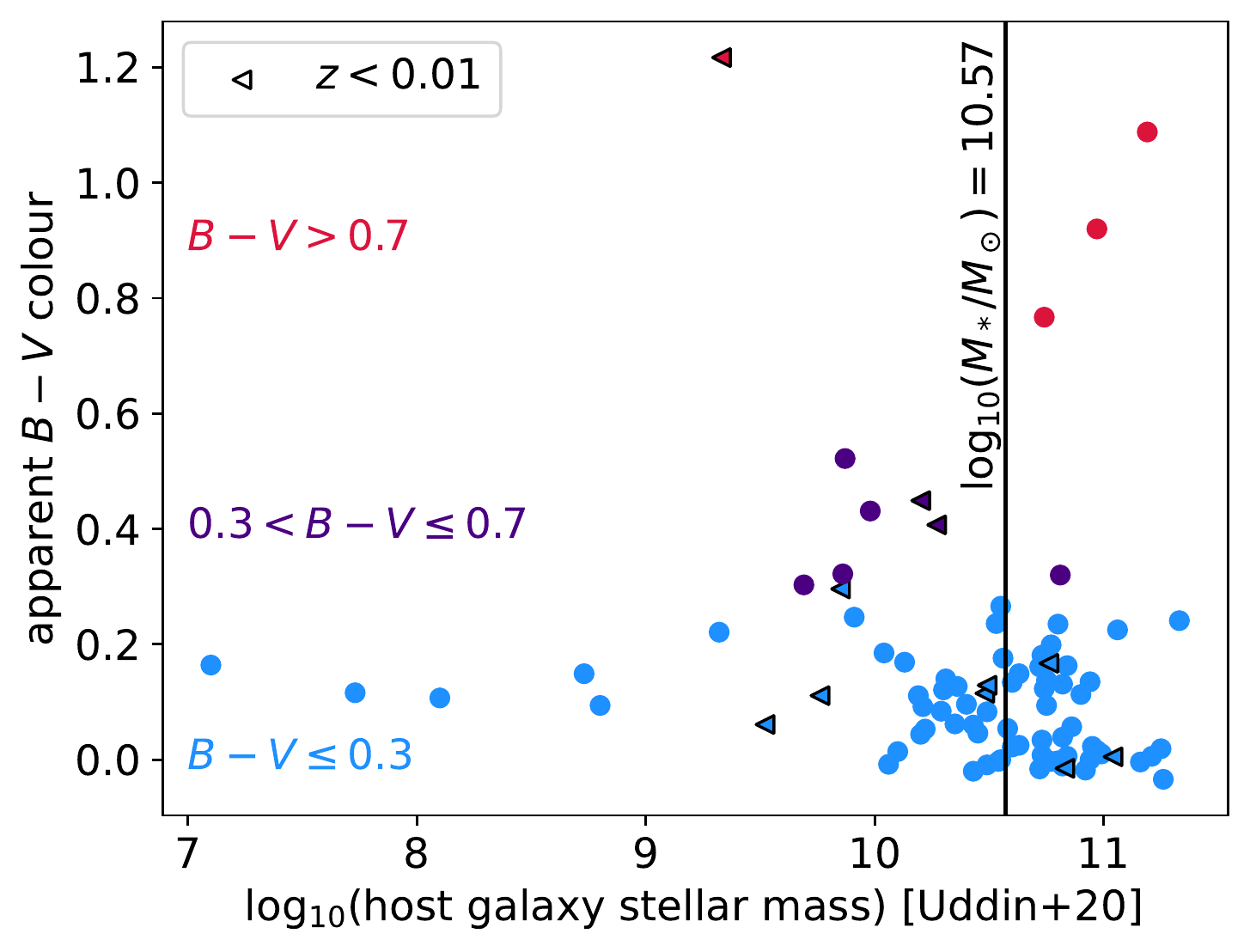}
    \caption{Apparent $B-V$ colours vs.\ host galaxy stellar mass for the 86 SNe Ia in our sample. Host mass estimates are from \citet{uddin20}. The median host galaxy stellar mass ($10^{10.57}~\mathrm{M}_\odot$) is marked with a vertical black line. Supernovae with an estimated $z_\text{CMB}<0.01$ are indicated as triangles with a black outlines.}
    \label{fig:colour_vs_mass}
\end{figure}

For all of these supernovae, we estimate values of \textsc{BayeSN}'s light curve shape parameter, $\theta_1$ (strongly correlated with optical decline rate; see section \ref{sec:bayesn} for definition), that are within the range spanned by the original \citet{mandel20} training set. Unlike for the \textsc{BayeSN} training set \citep[see][]{avelino19,mandel20}, we do not impose requirements on the temporal coverage of the light curves. We estimate that 75 out of the 86 SNe have apparent $B-V\leq0.3$ (consistent with the cut typically applied in cosmological analyses), but we do not exclude the redder supernovae at this stage. The sample covers a redshift range of $0.00365\leq z_\text{CMB}\leq0.08353$, with 11 SNe at $z_\text{CMB}<0.01$, and 15 SNe at $z_\text{CMB}>0.04$. The median host galaxy stellar mass of the sample \citep[using the masses from][]{uddin20} is $10^{10.57}~\mathrm{M}_\odot$. Figure \ref{fig:redshiftdist} shows the redshift distribution of the sample, broken down by apparent $B-V$ colour and host galaxy mass. From this, we see that the low- vs.\ high-mass split is reasonably even across the redshift range. However, the supernovae with redder apparent colours  ($B-V>0.3$) are concentrated towards the lower end on the redshift distribution. Figure \ref{fig:colour_vs_mass} shows a scatter plot of apparent $B-V$ colour vs.\ host mass. From this we can see that there is a fairly wide range of apparent colours on both sides of the median host mass ($10^{10.57}~\mathrm{M}_\odot$). The SNe with apparent $B-V\leq0.3$ are distributed fairly evenly across this mass split, with 36 SNe at lower masses, and 39 at higher masses. The middle apparent colour bin ($0.3<B-V\leq0.7$) leans towards the lower mass side, with $6/7$ SNe in this range having host masses less than the median. Conversely, the reddest SNe tend towards higher masses, with $3/4$ of these SNe being in hosts more massive than the median.

As well as our fiducial sample of 86, we also carry out a limited number of analyses (see Section \ref{sec:nirmax}) with a ``gold'' subset of 28 SNe Ia. These are selected as the 28 SNe out of our fiducial sample that overlap with the ``NIR@max'' cut of \citet{avelino19,mandel20} --  i.e.\ SNe with at least one NIR observation 2.5~d or more before the time of NIR maximum. This includes 28 out of the 30 CSP SNe from the \citet[][fig.\ 15]{mandel20} Hubble diagram -- two SNe from there (2007A and 2008bf) are not included in our fiducial sample here due to ambiguous host galaxies \citep{uddin20}. This ``gold'' subsample are all low reddening ($B-V<0.3$) by construction, and are all at $z<0.04$. Their median host galaxy mass is $10^{10.545}~\mathrm{M}_\odot$ -- very close to that of our full sample -- but we do not use this subset for any calculations with a mass split due to its small size. Nevertheless, dust constraints from this sample are of interest, as we would expect the residual intrinsic colour scatter to be smallest here (this can be seen in Fig.\ \ref{fig:colour_curves}).

In this paper, we will use the $BVriYJH$ photometry from \citet{krisciunas17}, for consistency with the passbands used in training our \textsc{BayeSN} SED model (see \S\ref{sec:fitting} or \citealp{mandel20} for more information about the model training). We do not include the CSP $u$- or $g$-band data in our fits. Our experience has shown that inclusion of the latter does not significantly influence our results, in the kind of analysis presented here, as the $g$-band tends to be somewhat redundant with the $B$- and $V$-band when all three of these bands are available. We did not include $u$/$U$-band data in the training of the current \textsc{BayeSN} model, which was restricted to $B$-band and longer wavelengths. Given the potential value of the UV to questions of dust, intrinsic diversity, and SN--host correlations \citep[see e.g.][]{brown10, foley12_uv, foley13, burns14, amanullah15, foley16, brown17, brown18, brown19}, its inclusion is a priority for future extensions of our model. Nevertheless, we demonstrate in Section \ref{sec:why_onir} that optical--NIR photometry already contains substantial information in this regard.

\section{Analysis Methods}
\label{sec:fitting}
We obtain our photometric distance estimates using the \citetalias{mandel20} version of the \textsc{BayeSN} SED model, described in \citet{mandel20}. This was trained on data from CSP-I \citep{krisciunas17}, the CfA/CfAIR Supernova Surveys \citep{jha99, hicken09, hicken12, woodvasey08, friedman15}, and elsewhere \citep{krisciunas03, krisciunas04a, krisciunas04b, krisciunas07, stanishev07, pignata08, leloudas09}, compiled in \citet{avelino19}. Our fits are carried out using Stan's \citep{carpenter17,stan21} standard Hamiltonian Monte-Carlo routine \citep{hoffman14,betancourt16} to sample the target posterior distributions defined mathematically in \S\ref{sec:jointpost}. To assess the quality of our MCMC chains, we use a number of standard diagnostics and procedures. We run multiple (4) independent chains in parallel starting from different initial locations in parameter space. We assess the mixing and convergence of the chains using the $\hat{R}$ (Gelman--Rubin) statistic and assess the effective sample size \citep{gelman92, vehtari19}, check that the estimated Bayesian fraction of missing information is not problematically low \citep{betancourt16_bfmi}, and confirm that our chains are free from divergent transitions \citep[see e.g.][]{betancourt14, betancourt15, betancourt17}. The final photometric distance estimate, and its error, for a given supernova is taken from the posterior mean and standard deviation of its distance modulus, marginalising over all other parameters. The full technical details of the \textsc{BayeSN} model and photometric distance estimation process are described in \citet{mandel20}, with a summary included below in Section \ref{sec:bayesn}.

\subsection{The \textsc{BayeSN} Model}
\label{sec:bayesn}
\textsc{BayeSN} models the rest-frame phase- and wavelength-varying host-dust-extinguished SED, $S_s(t,\lambda_r)$ of an SN Ia, $s$, via
\begin{multline}
    -2.5\log_{10}[S_s(t,\lambda_r)/S_0(t,\lambda_r)] = M_0 + W_0(t,\lambda_r)\\ + \delta M_s +\theta_1^sW_1(t,\lambda_r) + \epsilon_s(t,\lambda_r) +A_V^s\xi(\lambda_r;R_V^{(s)})
    \label{eq:SEDmodel}
\end{multline}
where $t$ is rest-frame phase (relative to time of $B$-band maximum) and $\lambda_r$ is rest-frame wavelength \citep[eq.\ 12]{mandel20}. The optical--NIR SN Ia spectral template, $S_0(t,\lambda_r)$, of \citet{hsiao07}, and a constant normalisation factor, $M_0\equiv-19.5$, are fixed a priori. The function $W_0(t, \lambda_r)$ warps the baseline spectral template to yield a mean intrinsic SED, whilst $W_1(t,\lambda_r)$ is a functional principal component (FPC) that captures the primary mode of intrinsic SED variation in the SN Ia population. Cubic spline representations $(\bm{W}_0,\bm{W}_1)$ of $W_0(t,\lambda_r)$ and $W_1(t,\lambda_r)$ are common to all supernovae, inferred during training, and fixed during photometric distance estimation. The $R_V$ parameter controlling the \citet{fitzpatrick99} dust law, $\xi(\lambda_r;R_V)$, was fully pooled (i.e.\ assumed to be the same for the entire sample) during training of the \citetalias{mandel20} model. We explore different treatments of this parameter at the photometric distance estimation stage, as described in Section \ref{sec:fittingconfigs}.

The latent parameters $A_V^s$, $\theta_1^s$, $\delta M_s$, $\epsilon_s(t,\lambda_r)$ take unique values for each supernova, $s$. These are, respectively, the host galaxy dust extinction in the $V$-band, $A_V^s$; a coefficient quantifying the effect of the $W_1(t,\lambda_r)$ FPC on the SED, $\theta_1^s$; a time- and wavelength-independent intrinsic magnitude offset, $\delta M_s$; and a time- and wavelength-varying function, $\epsilon_s(t,\lambda_r)$, representing residual variations in the intrinsic SED not captured by $\theta_1^sW_1(t,\lambda_r)$. This function models residual intrinsic colour variations across phase and wavelength beyond the effect of $\theta_1$. The population distributions of the latent parameters are modelled as:
\begingroup
\allowdisplaybreaks
\begin{align}
    A_V^s &\sim \exponential(\tau_A),\label{eq:avprior}\\
    \theta_1^s &\sim N(0,1),\\
    \delta M_s &\sim N(0,\sigma_0^2),\label{eq:delMprior}\\
    \bm{e}_s &\sim N(\bm{0},\bm{\Sigma}_\epsilon),\label{eq:epsilonprior}
\end{align}%
\endgroup
where $\bm{e}_s$ is a vector encoding a 2D cubic spline representation of the residual function $\epsilon(t,\lambda_r)$. The hyperparameters $\sigma_0$ and $\bm{\Sigma}_\epsilon$ are inferred during training, and fixed at the photometric distance estimation stage. The hyperparameter $\tau_A$ representing the population mean of the dust extinction $A_V$ is ordinarily fixed during photometric distance estimation (when supernovae are treated independently), but can also be inferred if the full sample is fit jointly (see Section \ref{sec:fittingconfigs}~\ref{itm:pp}).

In our generative model for SN Ia photometry, the host-dust-extinguished SED, $S_s(t,\lambda_r)$, is dimmed by the distance modulus, $\mu_s$, redshifted, and extinguished by Milky Way dust (using the \citealp{fitzpatrick99} dust law, and \citealp{schlafly11} reddening map). The resulting observer frame SED, now reddened by both host and Milky Way dust, is then integrated through photometric bandpasses to yield model photometry. This process mathematically defines a likelihood function that is used to compare the model to observed photometry. During training, this likelihood forms the outer layer of a hierarchical model under which all population-level $(\bm{W}_0, \bm{W}_1, \bm{\Sigma}_\epsilon, \sigma_0, \tau_A, R_V)$ and supernova-level parameters $(A_V^s, \theta_1^s, \delta M_s, \bm{e}_s, \mu_s)$ are inferred simultaneously for a sample of SNe Ia (see \citealp{mandel20} for complete specification). The latter parameters are marginalised over, to yield posterior estimates of the former that are carried forward into photometric distance estimation. During photometric distance estimation, population-level parameters are ordinarily kept fixed, and external constraints on distance (from, e.g., redshift + an assumed cosmology, or a redshift-independent distance probe) are relaxed. Under this fitting mode, all supernovae are conditionally independent in the posterior (see \citealp{mandel20}). In this work, we conduct several variations of the photometric distance estimation process, as detailed in Section \ref{sec:fittingconfigs}. For some of these, we include and marginalise over population level parameters (e.g.\ $R_V$, or hyperparameters defining its population distribution) during photometric distance estimation. In these cases, we sample from the joint posterior over all supernovae. Section \ref{sec:jointpost} discusses this.

\subsection{Fitting Configurations}
\label{sec:fittingconfigs}
We take a number of different approaches to our light curve fits, yielding several sets of photometric distance estimates for our sample. Our primary fitting configurations are:
\begin{enumerate}
    \item \label{itm:m20} \emph{Fixed $R_V$}: Fitting all available optical-only ($BVri$), NIR-only ($YJH$), or optical+NIR ($BVriYJH$) data with the default \citetalias{mandel20} \textsc{BayeSN} model trained by \citet{mandel20}. In this fitting mode, each supernova is fitted independently from the others, with a fixed $R_V=2.89$ \citep[the global $R_V$ inferred by][]{mandel20} assumed.
    \item \label{itm:pp} \emph{Population $R_V$}: Fitting all $BVri$ or $BVriYJH$ data using the \citetalias{mandel20} model and \emph{partial pooling} of $R_V$. This means that each supernova $s$ has its own individual $R_V^s$, drawn from a common population distribution described by hyperparameters that are estimated from the data\footnote{In contrast, \emph{complete pooling} would imply a single common $R_V$ for all supernovae, whilst \emph{no pooling} would remove the assumption of a common $R_V$, or population distribution thereof, and would imply that all supernovae are fit completely independently. For a review of these distinctions, consult \citet[ch.~5]{gelman14}.}. Under this configuration, we fit all 86 SNe Ia in our CSP-I sample jointly. \textsc{BayeSN} hyperparameters defining the intrinsic SED $(\bm{W}_0, \bm{W}_1, \bm{\Sigma}_\epsilon, \sigma_0)$ are kept fixed to the \citetalias{mandel20} estimates. The hyperparameter, $\tau_A$, specifying the mean of the $A_V$ population distribution (Eq.\ \ref{eq:avprior}) is inferred from the sample, along with the parameters $(\mu_R,\sigma_R)$ specifying a truncated normal\footnote{We define $X\sim\text{Trunc-}N(\mu,\sigma^2,a,b)$ as a random variable drawn from a normal distribution with mean, $\mu$, and standard deviation, $\sigma$, with lower and upper truncation at $a$ and $b$, respectively. The probability density function for $a\leq X\leq b$ is \begin{equation*}P(X|\mu,\sigma,a,b)=\frac{1}{\sigma}\frac{\phi(\xi)}{\Phi(\beta)-\Phi(\alpha)},\end{equation*} and is zero for $X < a$, $X > b$. Here, $\xi=(X-\mu)/\sigma$, $\alpha=(a-\mu)/\sigma$, $\beta=(b-\mu)/\sigma$, and $\phi(z)$ and $\Phi(z)$ are respectively the PDF and CDF of a standard normal random variable $z$.} $R_V$ population distribution,
    \begin{equation}
        R_V^s \sim \text{Trunc-}N(\mu_R, \sigma_R^2, a=0.5, b=\infty).
        \label{eq:rvprior}
    \end{equation}
    Hyperpriors on $(\tau_A,\mu_R,\sigma_R)$ are as specified in \citet{mandel20,thorp21}. As in \citet{thorp21}, our default $\sigma_R$ hyperprior is a half-normal, $\sigma_R\sim \text{Half-}N(0,2^2)$. This is close to flat across all possible $\sigma_R$ values of practical interest (see Fig.\ \ref{fig:distributions}), but avoids the hard boundary associated with a uniform prior over a finite range. We consider our sensitivity to this choice in Appendix \ref{app:sigmaRprior}, where we also try $\sigma_R\sim\text{Half-}N(0,1^2)$, $\sigma_R\sim U(0,4)$, and an improper flat prior on positive $\sigma_R$, and a hierarchical inverse-$\chi^2$ prior \citep[inspired by][]{burns14}. We also explore two alternative forms for the $R_V$ population distribution (a skew-normal or Student's $t$) in Appendix \ref{app:altdistributions}. The approach of partially pooling $R_V$ allows for principled sharing of information between SNe. This leads to more reliable inference of the population distribution of $R_V$ than one would get from a collection of individual $R_V$ estimates obtained with no pooling. Our fully Bayesian approach also enables us to estimate distances whilst marginalising over the individual $R_V^s$ and $A_V^s$ of the sample, as well the hyperparameters of their population distributions -- something previous mass step analyses have not been able to do.
    \item \label{itm:binned} \emph{Binned population $R_V$}: A variation of the population $R_V$ configuration \ref{itm:pp} where the sample of 86 SNe Ia is subdivided by either SN Ia apparent $B-V$ colour \citep[\`a la][]{mandel11,burns14}, or host galaxy stellar mass \citep[\`a la][]{brout20,thorp21}. In each bin, an independent $R_V$ population distribution is assumed, with the same form as Equation \ref{eq:rvprior}. When binning by host galaxy stellar mass, we use the mass estimates from \citet{uddin20}, and divide the sample at either $10^{10}~\mathrm{M}_\odot$ (a conventional choice since \citealp{sullivan10}; see discussion in \citealp{ponder20}, \S5.7), or $10^{10.57}~\mathrm{M}_\odot$ (the median host mass for our sample of 86 SNe). When binning by apparent $B-V$ colour\footnote{Strictly speaking, we compute host-dust-extinguished rest-frame $B-V$ colour at the time of $B$-band maximum. This quantity is fairly ``close to the data'' and is quite insensitive to the exact fitting configuration used.}, we use our own estimates of this quantity from fits in the fixed $R_V$ configuration \ref{itm:m20}. We adopt either a 2-bin configuration, splitting at $B-V=0.3$ (roughly equivalent to a typical $c_\text{SALT2}=0.3$ cosmology cut), or a 3-bin configuration with bins of $B-V\leq0.3$, $0.3<B-V\leq0.7$, and $B-V>0.7$.
    \item \label{itm:bs21} \emph{Free $R_V$, \citetalias{brout20} prior}: Fitting all $BVri$ or $BVriYJH$ data using the \citetalias{mandel20} model, with a free $R_V$ for each supernova, and an $R_V$ prior based on the \citep[\citetalias{brout20}]{brout20} model:
    \begin{equation}
        R_V^s \sim \begin{cases}
            \text{Trunc-}N(\mu_R=1.50, \sigma_R^2=1.3^2) \text{ if } M_*\geq10^{10}~\mathrm{M}_\odot\\
            \text{Trunc-}N(\mu_R=2.75, \sigma_R^2=1.3^2) \text{ if } M_*<10^{10}~\mathrm{M}_\odot
        \end{cases}.
        \label{eq:bs21rvprior}
    \end{equation}
    As in Eq.\ \ref{eq:rvprior} (and consistent with \citealp{brout20}), we adopt a lower $R_V$ truncation of $a=0.5$, with no upper truncation (i.e.\ $b=\infty$).
    \item \label{itm:u} \emph{Free $R_V$, uniform prior}: Fitting all $BVri$ or $BVriYJH$ data using the \citetalias{mandel20} model, with free $R_V$ for each supernova and a uniform prior, $R_V^s\sim U(1,6)$. The lower limit here is chosen to be around the theoretical minimum value of $R_V$ \citep[from the Rayleigh scattering limit, see][]{draine03}. The upper limit is set to encompass the highest $R_V$ values reported \citep[see e.g.][]{cardelli89, fitzpatrick99, fitzpatrick07, fitzpatrick09} along sight lines in the Milky Way (although such extreme values are found to be vanishingly rare in our Galaxy -- see \citealp{schlafly16}).
\end{enumerate}

\subsection{Posterior Distributions}
\label{sec:jointpost}
In the standard \textsc{BayeSN} photometric distance estimation procedure \citep[see][\S2.8]{mandel20}, all population-level parameters of the model are fixed to their training values, and every supernova can be treated independently. In this case, the posterior distribution is given by Eq. 30 in \citet{mandel20}. Our \emph{fixed $R_V$} fitting configuration (Section \ref{sec:fittingconfigs}~\ref{itm:m20}) is exactly equivalent to this. Our \emph{free $R_V$} configurations are very similar, with every supernova being independent, albeit with its host galaxy $R_V^s$ as a free parameter. In this case, the posterior distribution we sample for a supernova, $s$, can be written as
\begin{multline}
    P(\mu_s, \bm{\phi}_s| \bm{\hat{f}}_s, \bm{\hat{W}}_0, \bm{\hat{W}}_1, \bm{\hat{\Sigma}}_\epsilon, \hat{\sigma}_0, \hat{\tau}_A) \\ \propto P(\bm{\hat{f}}_s | \mu_s, A_V^s, R_V^s, \theta_1^s, \delta_M^s, \bm{e}_s, \bm{\hat{W}}_0, \bm{\hat{W}}_1) \times P(\mu_s) \\ \times P(A_V^s|\hat{\tau}_A)\times P(R_V^s) \times P(\theta_1^s)\times P(\bm{e}_s|\bm{\hat{\Sigma}}_\epsilon)\times P(\delta M_s|\hat{\sigma}_0),
    \label{eq:individpost}
\end{multline}
where $(\bm{\hat{W}}_0, \bm{\hat{W}}_1, \bm{\hat{\Sigma}}_\epsilon, \hat{\sigma}_0, \hat{\tau}_A)$ are population-level hyperparameter estimates from training\footnote{We use the \citetalias{mandel20} values of these, available at: \url{https://github.com/bayesn/bayesn-model-files}}, $\bm{\phi}_s=(\mu_s, A_V^s, R_V^s, \theta_1^s, \delta M_s, \bm{e}_s)$ are supernova-level latent parameters, and $\bm{\hat{f}}_s$ are the observed fluxes for supernova $s$. The first factor on the right-hand side is the light curve data likelihood (c.f. \S 2.1 of \citealp{mandel20}), while the factors on the third line are population distributions. When estimating a photometric distance, $P(\mu_s)\propto1$. Otherwise, it is based on the external distance constraint (see Section \ref{sec:resids} for details of these). For our \emph{free $R_V$} configuration with a uniform prior, $P(R_V^s)=U(R_V^s|1,6)$. When using the \citetalias{brout20} prior, $P(R_V^s)$ is dependent on the host galaxy mass of supernova $s$, and follows from Equation \ref{eq:bs21rvprior}.

Equation \ref{eq:individpost} shows the construction of the posterior distribution for our \emph{free $R_V$} fitting configurations where no pooling takes place (described in Section \ref{sec:fittingconfigs}~\ref{itm:bs21} and \ref{itm:u}). In our \emph{population $R_V$} fitting configurations (described in Section \ref{sec:fittingconfigs}~\ref{itm:pp} and \ref{itm:binned}), we must define a joint posterior over all supernovae. For the \emph{population $R_V$} configuration without binning by mass or colour, we have
\begin{multline}
    P(\{\mu_s, \bm{\phi}_s\}, \mu_R, \sigma_R, \tau_A| \{\bm{\hat{f}}_s\}, \bm{\hat{W}}_0, \bm{\hat{W}}_1, \bm{\hat{\Sigma}}_\epsilon, \hat{\sigma}_0) \\\propto \bigg[\prod_s P(\bm{\hat{f}}_s | \mu_s, A_V^s, R_V^s, \theta_1^s, \delta_M^s, \bm{e}_s, \bm{\hat{W}}_0, \bm{\hat{W}}_1) \\ \times P(\mu_s) \times P(A_V^s|\tau_A)\times P(R_V^s|\mu_R,\sigma_R) \times P(\theta_1^s) \\ \times P(\bm{e}_s|\bm{\hat{\Sigma}}_\epsilon)\times P(\delta M_s|\hat{\sigma}_0)\bigg] \times P(\tau_A) \times P(\mu_R) \times P(\sigma_R).
    \label{eq:jointpost}
\end{multline}
Our hyperpriors, $P(\mu_R)$, $P(\sigma_R)$, $P(\tau_A)$, are as specified in \citet{mandel20}, \citet{thorp21}, and Section \ref{sec:fittingconfigs}~\ref{itm:pp}. 

For the binned population $R_V$ fitting configurations (Section \ref{sec:fittingconfigs}~\ref{itm:binned}), some or all of the population-level hyperparameters $(\mu_R, \sigma_R, \tau_A)$ are split by host mass or apparent colour and Eq. \ref{eq:jointpost} is modified accordingly. When binning by mass, we assume that all three of these hyperparameters are split for the two mass bins. When binning by apparent colour, we maintain a common $A_V$ population distribution across the colour bins (i.e.\ an exponential distribution parameterised by a single common $\tau_A$).

\subsection{Hubble Residuals}
\label{sec:resids}
From each set of photometric distance moduli, we compute Hubble residuals by subtracting off external distance modulus estimates computed from the supernova host galaxy redshifts under a fiducial cosmology of $H_0 = 73.24~\text{km\,s}^{-1}\,\text{Mpc}^{-1}$, $\Omega_M = 0.28$, and $\Omega_\Lambda=0.72$ \citep{riess16}. The redshifts are corrected to the CMB rest frame and for local flows using peculiar velocities estimated using the \citet{carrick15} flow model\footnote{\url{https://cosmicflows.iap.fr/}}. We compute the uncertainty on an external distance estimate as in \citet[eq.\ 8]{avelino19}, assuming a peculiar velocity uncertainty of 150~km\,s$^{-1}$ \citep{carrick15}. Where redshift-independent distance constraints are available for nearby SNe \citep[e.g.\ from Cepheid variables, Tully--Fisher, or similar, as listed in][table 4]{avelino19}, we use these (and their uncertainties) instead. Uncertainties on the external distance estimates are added in quadrature to the posterior uncertainties on the \textsc{BayeSN} photometric distances to obtain the uncertainties on the Hubble residuals. The posterior uncertainties on the photometric distances naturally include the intrinsic scatter via marginalisation of the above posteriors.

\subsection{Searching for a Mass Step}
\label{sec:searching}
We use the best-fit estimates of host galaxy stellar mass reported in \citet[table C1]{uddin20}. For a given set of Hubble residuals, we estimate the step size at a given mass by taking the difference of the weighted means of the Hubble residuals either side of the chosen step. The weight of the Hubble residual is the inverse of the sum of the squares of its photometric distance uncertainty and external distance uncertainty. The quadrature sum of the errors on the weighted means gives the uncertainty in step size. We compute the step sizes at $10^{10}~\mathrm{M}_\odot$, and $10^{10.57}~\mathrm{M}_\odot$ (the median host mass for our sample of 86 SNe Ia). 

We also compute the maximum likelihood estimate (MLE) of the step location for each set of Hubble residuals, and estimate the step size here. We do this following the methodology described in \citet[appendix A]{thorp21}, which finds the step location that maximises $P(\text{Hubble resids.}|\text{step loc.})$, having marginalised over the the residual Hubble diagram scatter and the mean Hubble residual on either side of the step. To avoid spurious maxima, we restrict the maximisation to step locations that fall within the interquartile range of the host galaxy masses of our sample ($\approx10^{10.20}$--$10^{10.82}~\mathrm{M}_\odot$).

\section{Why Optical and NIR Data Are Needed}
\label{sec:why_onir}
Several past analyses have shown and exploited the fact that optical and NIR data combined can be used to obtain improved constraints on SN Ia host galaxy dust \citep[e.g.][]{krisciunas07, folatelli10, mandel11, mandel20, burns14, johansson21}. The unique advantage gained by adding NIR data to optical comes from the additional colour information provided -- with a wider wavelength range, it becomes possible to disentangle total extinction (e.g.\ $A_B$ or $A_V$) from $R_V$ using colour information alone. When only optical colours are available, additional information is needed to obtain robust estimates (e.g.\ information about absolute magnitude, derived from an independent estimate of the SN Ia distance).

\begin{figure*}
    \centering
    \includegraphics[width=\linewidth]{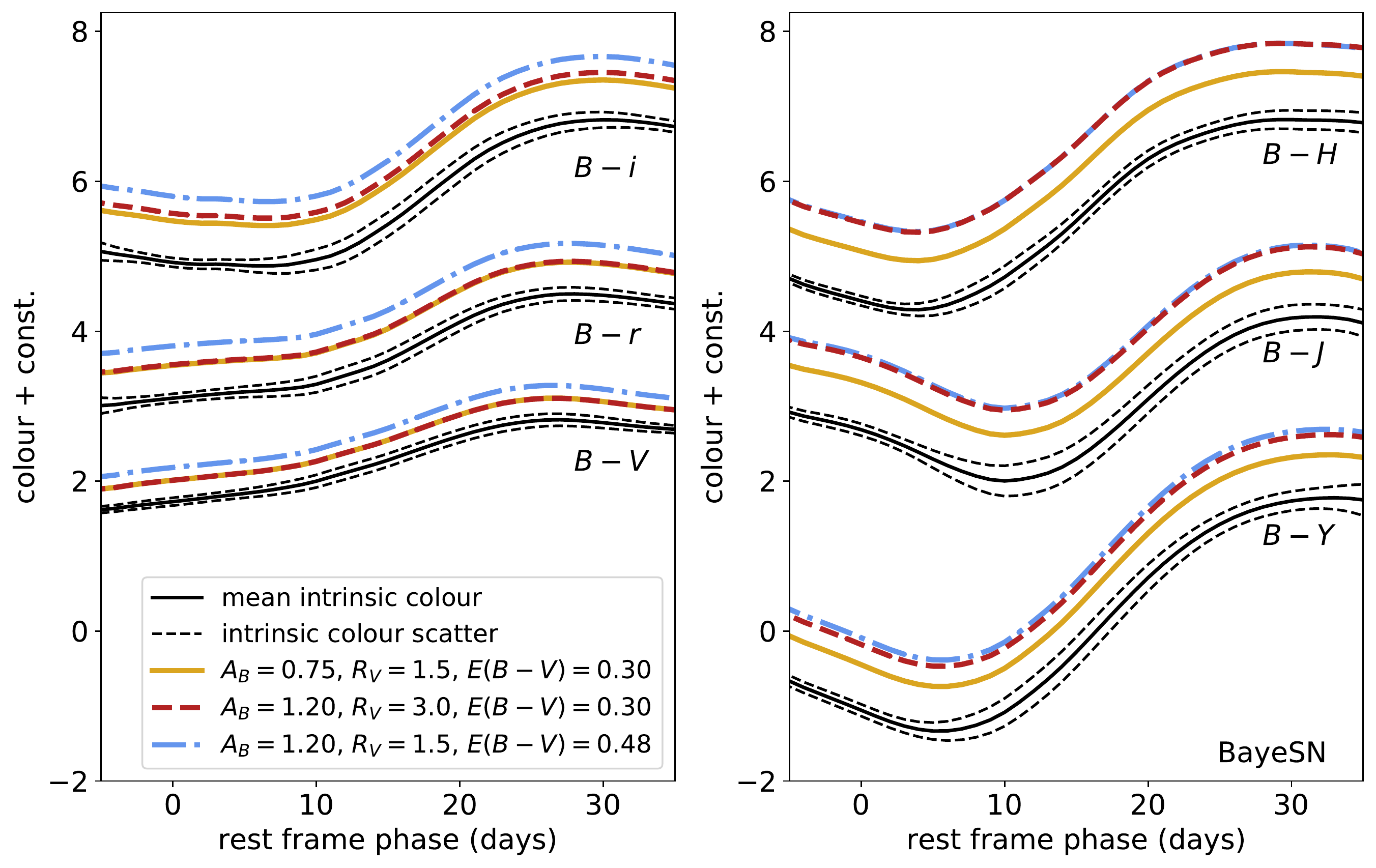}
    \caption{Colour curves in the rest-frame CSP passbands simulated using the \citetalias{mandel20} \textsc{BayeSN} model. Solid and dashed black lines show the mean intrinsic colour and intrinsic colour scatter ($\pm 1\sigma$), after correcting the light curve shape to the population mean $\theta_1$. Coloured lines show how dust affects the mean colour curves. The yellow and red lines correspond to different dust configurations which give rise to $E(B-V)=0.3$~mag. The blue lines correspond to a higher reddening of $E(B-V)=0.48$~mag, but the same $R_V$ as the yellow lines. Vertical offsets are (1.75, 3.25, 5.75, 0.0, 3.5, 5.5) for ($B-V$, $B-r$, $B-i$, $B-Y$, $B-J$, $B-H$).}
    \label{fig:colour_curves}
\end{figure*}

To see why optical and NIR data are so powerful when used in combination, we can look at the effect of dust on SN Ia colour curves for a range of optical and optical$-$NIR colours. Figure \ref{fig:colour_curves} shows $B-X$ colour curves (for $X\in VriYJH$), simulated using the \citetalias{mandel20} \textsc{BayeSN} model. In Figure \ref{fig:colour_curves}, the black lines show the distribution of intrinsic colour curves, after correction for light curve shape. The yellow ($A_B=0.75$, $A_V=0.45$, $R_V=1.5$) and red ($A_B=1.2$, $A_V=0.9$, $R_V=3.0$) lines show two different combinations of reddening law and $B$-band extinction that yield the same reddening: $E(B-V)=0.3$. The blue line ($A_B=1.2$, $A_V=0.72$, $R_V=1.5$) shows a third dust configuration, with the same $B$-band extinction as the red line, the same dust law as the yellow line, but a higher reddening: $E(B-V)=0.48$. From the closeness of the red and blue curves in the right hand panel of Figure \ref{fig:colour_curves}, one can immediately see that the optical$-$NIR colours (i.e.\ $B-J$, $B-H$) are highly sensitive to the total $B$-band extinction, but almost unaffected by $R_V$. Conversely, the optical colour curves (i.e.\ $B-V$, $B-r$) are mostly sensitive to the reddening, $E(B-V)$. This can be seen from the closeness of the red and yellow curves in the left hand panel of Figure \ref{fig:colour_curves}. 

Consequently, by constraining $E(B-V)$ using optical colours, and extinction (e.g.\ $A_B$) using optical$-$NIR colours, one can imagine directly estimating $R_V=A_B/E(B-V) - 1$. In practice, we do not directly estimate $R_V$ from inferring extinction and reddening derived from colours in this way. Instead, our \textsc{BayeSN} hierarchical model (see Section \ref{sec:bayesn}) implicitly leverages the colour information from all available passbands at all observed times, weighting the data appropriately based on the embedded residual intrinsic scatter model to yield well motivated uncertainties on the dust extinction and dust law $R_V$.

\section{Why Hierarchical Bayes is Needed}
\label{sec:why_hb}
To motivate our hierarchical Bayesian methodology, we present here a simulation-based exercise illustrating the challenges of correctly estimating the $R_V$ population distribution from a sample of supernovae. To do this, we simulate several sets of CSP-like light curves from the \citetalias{mandel20} \textsc{BayeSN} forward generative model. For each simulation, we generate 86 light curves with the same redshifts, light curve shapes ($\theta_1^s$), host galaxy dust extinction ($A_V^s$), and observation patterns (observation times and passbands) as the real supernovae in our fiducial CSP sample. The intrinsic residuals ($\delta M_s$ and $\epsilon_s(t,\lambda_r$)) are randomly realised for each simulated supernova from the population distributions inferred during training of the \citetalias{mandel20} model. We generate three variations of this simulation with different input $R_V$ distributions, all with population mean $\mu_R=2.7$, and with population standard deviations of  $\sigma_R=0.0$, 0.5, and 1.0. For each simulation we generate random realisations of the true $\{ R_V^s\}$.

We perform several sets of fits to each of these simulated samples, using the \citetalias{mandel20} \textsc{BayeSN} model. First, we fit all of the light curves independently, assuming a uniform prior on each individual $R_V^s$ -- this corresponds to the free $R_V$ fitting configuration, as described in Section \ref{sec:fittingconfigs}~\ref{itm:u}, although here our exact prior is $R_V^s\sim U(0.5,6.0)$\footnote{This is chosen for consistency with the simulations, where a lower truncation of $R_V^s>0.5$ was applied to the Gaussian population distributions}. We then repeat these individual light curve fits, but with the \textsc{BayeSN} residual intrinsic scatter model switched off during fitting (i.e. $\delta M_s$ and $\bm{e}_s$ are fixed to zero). This forces all apparent colour variation to be explained by a combination of the principal SED shape parameter ($\theta_1^s$), and dust extinction ($A_V^s$, $R_V^s$), and ignores any possible variation arising from residual intrinsic colour scatter. This should be somewhat analogous to the \textsc{SNooPy} \texttt{color\_model} \citep{burns11, burns14}. From each of these fits, one might take the posterior means, $\hat{R}_V^s$, and standard deviations, $\hat{\sigma}_s$, as point estimates of each supernova's $R_V^s$. We then use these to perform two different estimates of the parameters ($\mu_R$, $\sigma_R$) population's $R_V$ distribution:
\begin{enumerate}
    \renewcommand{\theenumi}{(\Alph{enumi})}
    \item \label{itm:stddev} \emph{Naive}: The simple sample mean and sample standard deviation of $\{\hat{R}_V^s\}$ are taken as estimates ($\hat{\mu}_R$, $\hat{\sigma}_R$) of the population mean and standard deviation, respectively.
    \item \label{itm:shrinkage} \emph{Shrinkage}: We perform a shrinkage estimate of ($\mu_R$, $\sigma_R$) by assuming a simple normal--normal model for our set of point estimates, $\{\hat{R}_V^s,\hat{\sigma}_s\}$. The posterior is
    \begin{equation}
        P(\mu_R,\sigma_R|\{\hat{R}_V^s\}) \propto P(\mu_R,\sigma_R)\prod_{s=1}^{N_\text{SN}} N(\hat{R}_V^s|\mu_R,\hat{\sigma}_s^2+\sigma_R^2)\label{eq:8sch},
    \end{equation}
    as in \citet[eq.~5.18]{gelman14}. We maximize Eq.~\ref{eq:8sch}, assuming a flat prior on the parameters -- i.e.\ $P(\mu_R,\sigma_R)\propto1$ -- to obtain a MAP estimate of $\mu_R$, $\sigma_R$.
\end{enumerate}
Finally, we compare our population distribution estimates from methods \ref{itm:stddev} and \ref{itm:shrinkage} to our preferred approach:
\begin{enumerate}
    \setcounter{enumi}{2}
    \renewcommand{\theenumi}{(\Alph{enumi})}
    \item \label{itm:full} \emph{Full hierarchical Bayes (H.B.)}: Estimating $\mu_R$ and $\sigma_R$ from a hierarchical Bayesian analysis of all light curves simultaneously, under an $R_V$ population model. This is equivalent to the population $R_V$ fitting configuration with partial pooling in Section \ref{sec:fittingconfigs}~\ref{itm:pp}.
\end{enumerate}

We expect the full hierarchical Bayesian method \ref{itm:full} to give the most reliable estimates of the population distribution. When the fits are carried out with the residual scatter model included, we would expect the shrinkage estimator \ref{itm:shrinkage} to provide a good  estimate of the population standard deviation, $\sigma_R$, but the naive approach \ref{itm:stddev} to overestimate the $R_V$ dispersion as it does not account for the extra scatter induced by the uncertain individual $\hat{R}_V^s$ estimates. When the fits are carried out with the residual scatter model turned off, we expect the uncertainties on the individual $\hat{R}_V^s$ estimates to be underestimated, \textit{and} these estimates to be over-dispersed (since the apparent colours of all supernovae with a given light curve shape are forced to map back to the same intrinsic colours, e.g. the solid black line in Fig.~\ref{fig:colour_curves}). We therefore expect the shrinkage estimator \ref{itm:shrinkage} to overestimate $\sigma_R$ in this case, as the widely dispersed and over-confident $\hat{R}_V^s$ estimates will cause the required level of shrinkage to be underestimated. We expect the naive method \ref{itm:stddev} to perform even worse.

\begin{table}
    \centering
    \begin{threeparttable}
        \caption{Inferred $R_V$ population distribution parameters on simulated datasets.}
        \label{tab:sim_recovery}
        \begin{tabular}{c c c c c c}\toprule
            \multicolumn{2}{c}{Input\tnote{a}} & Method\tnote{b} & Res. & \multicolumn{2}{c}{Estimated\tnote{d}}\\
            \cmidrule(r){1-2} \cmidrule(l){5-6}
            $\mu_R$ & $\sigma_R$ & & Scatter\tnote{c} & $\hat{\mu}_R$ & $\hat{\sigma}_R$\\\midrule
            2.7 & 0.0 & \emph{Naive} \ref{itm:stddev} & No & 3.21 & 1.60 \\
            & & & Yes & 3.18 & 0.63 \\
            & & \emph{Shrink.}~\ref{itm:shrinkage} & No & 3.18 & 1.61 \\
            & & & Yes & 2.70 & 0.05 \\
            & & \emph{H.B.}~\ref{itm:full} & Yes & $2.67\pm0.06$ & 0.12~(0.20) \\
            2.7 & 0.5 & \emph{Naive} \ref{itm:stddev} & No & 3.16 & 1.66 \\
            & & & Yes & 3.18 & 0.81 \\
            & & \emph{Shrink.}~\ref{itm:shrinkage} & No & 3.14 & 1.67 \\
            & & &  Yes & 2.91 & 0.60 \\
            & & \emph{H.B.}~\ref{itm:full} & Yes & $2.80\pm0.13$ & $0.63\pm0.12$ \\
            2.7 & 1.0 & \emph{Naive} \ref{itm:stddev} & No & 3.19 & 1.73\\
            & & & Yes & 3.09 & 0.97 \\
            & & \emph{Shrink.}~\ref{itm:shrinkage} & No & 3.17 & 1.73 \\
            & & & Yes & 2.82 & 0.88 \\
            & & \emph{H.B.}~\ref{itm:full} & Yes & $2.66\pm0.18$ & $1.05\pm0.19$\\
            \bottomrule
        \end{tabular}
        \begin{tablenotes}
            \item [a] Input values of the true $R_V$ population parameters.
            \item [b] Method used to estimate parameters (see text in \S\ref{sec:why_hb}).
            \item [c] If Yes, \textsc{BayeSN}'s residual intrinsic scatter model was included and marginalised over when fitting light curves.
            \item [d] For the full H.B.\ method \ref{itm:full}, reported estimates are either posterior mean $\pm$ std.\ dev.\ or 68~(95)~\% upper bounds.
        \end{tablenotes}
    \end{threeparttable}
\end{table}

Table \ref{tab:sim_recovery} reports the full set of results from our simulation-based tests. For the naive method \ref{itm:stddev}, the estimate of the population mean, $\hat{\mu}_R$, is consistently biased towards the mean of the individual $R_V^s$ prior used in the fits, i.e. $\mathbb{E}(R_V^s) = 3.25$ for $R_V^s\sim U(0.5,6.0)$. When analysing the individual fits results with the shrinkage estimator \ref{itm:shrinkage}, this remains true for fits where residual scatter was ignored. However, for the fits where residual scatter was accounted for, the population mean estimate from method \ref{itm:shrinkage} is much closer to the truth. From the fully hierarchical analysis \ref{itm:full}, the $\hat{\mu}_R$ estimates are within one posterior standard deviation of the truth for all three simulations.

\begin{figure}
    \centering
    \includegraphics[width=\linewidth]{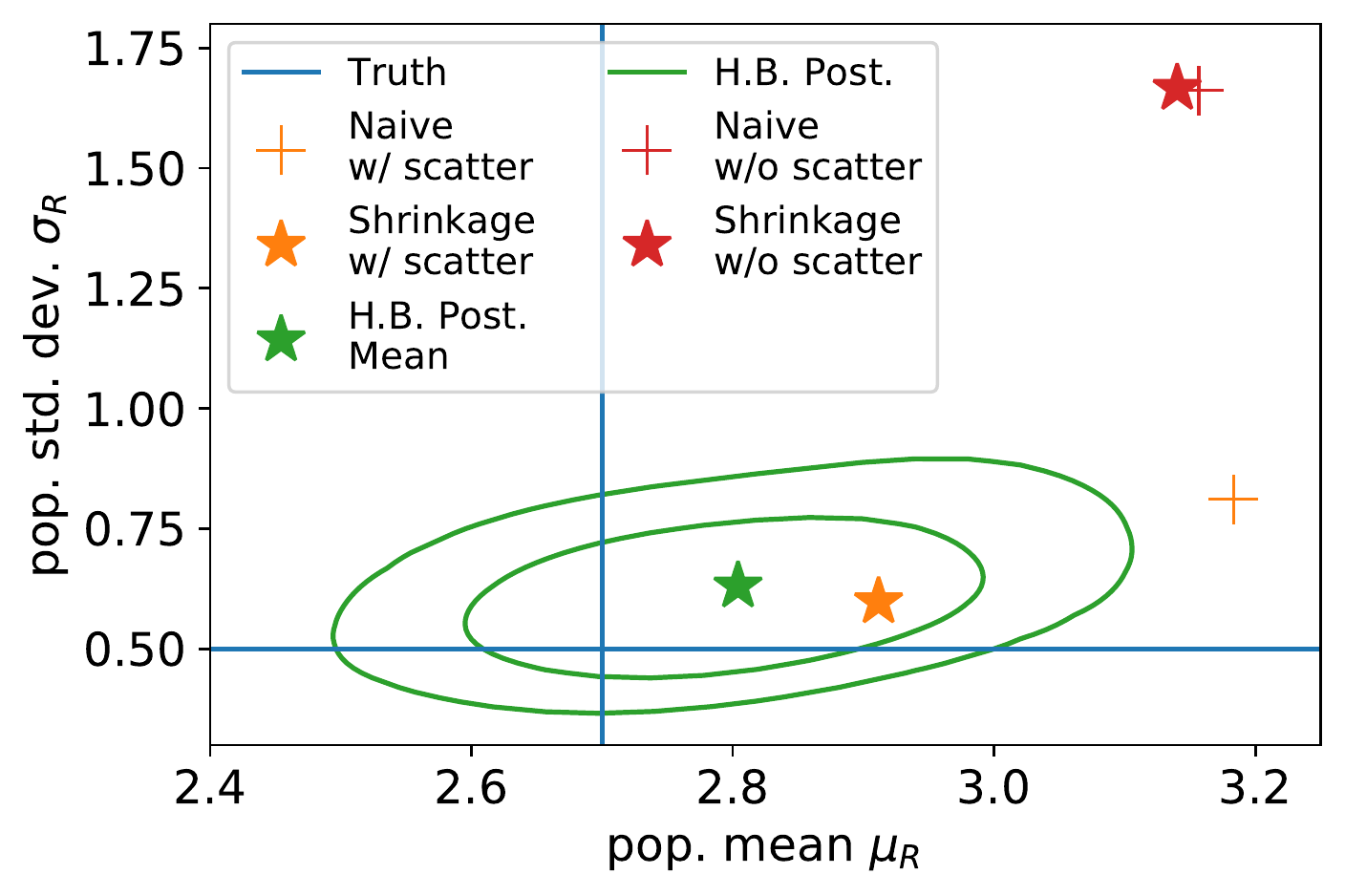}
    \caption{Visual comparison of different approaches for estimating the $R_V$ population distribution hyperparameters of a simulated population of 86 SNe Ia. The simulation was generated with a population mean of $\mu_R=2.7$, and population standard deviation of $\sigma_R=0.5$ (indicated by the blue crosshairs). The green contours show the 2D marginal posterior distribution (68 and 95 per cent credible intervals) obtained from a full hierarchical Bayesian analysis of the simulated data, with the green star showing the posterior mean for each hyperparameter. The orange and red crosses show the naive sample mean and standard deviation of the $\{R_V^s\}$ point estimates made by fitting each supernova individually with \textsc{BayeSN}, both with (orange cross) and without (red cross) \textsc{BayeSN}'s residual scatter model turned on. Orange and red stars show the result of applying a simple shrinkage estimator to the same sets of $R_V^s$ point estimates. Numerical results are summarised in Table \ref{tab:sim_recovery}.}
    \label{fig:sim_recovery}
\end{figure}

The estimates of the population standard deviation conform well to our expectations. When the residual scatter is ignored during the fits, the estimate of the population standard deviation, $\hat{\sigma}_R$ is poor when using the naive method \ref{itm:stddev} or shrinkage estimator \ref{itm:shrinkage}. For all three simulations, the estimate is close to the standard deviation one would expect from the individual $R_V^s$ prior used in the fits ($\text{Std.\ Dev.}(R_V^s) = 1.59$ for $R_V^s\sim U(0.5,6.0)$). This reflects the poor estimates of the individuals' $R_V^s$ when residual scatter is neglected. When residual scatter is included in the individual fits, our $\hat{\sigma}_R$ estimate is still biased high for the two simulations with narrower population distributions when using the naive method \ref{itm:stddev}. When using the shrinkage estimator \ref{itm:shrinkage}, the $\hat{\sigma}_R$ estimates are close to the truth for all three simulations. The fully hierarchical analysis \ref{itm:full} performs well, as expected. For the two simulations with non-zero input population variance, the recovered $\hat{\sigma}_R$ values are consistent with the truth. For the simulation with input $\sigma_R=0$, the posterior from the full hierarchical Bayesian approach peaks at zero, and places a 95 per cent upper bound of 0.20. Figure \ref{fig:sim_recovery} shows a comparison of the different methods for estimating the population mean and standard deviation, for the simulation with input parameters $\mu_R=2.7$ and $\sigma_R=0.5$.

\begin{figure*}
    \centering
    \includegraphics[width=\textwidth]{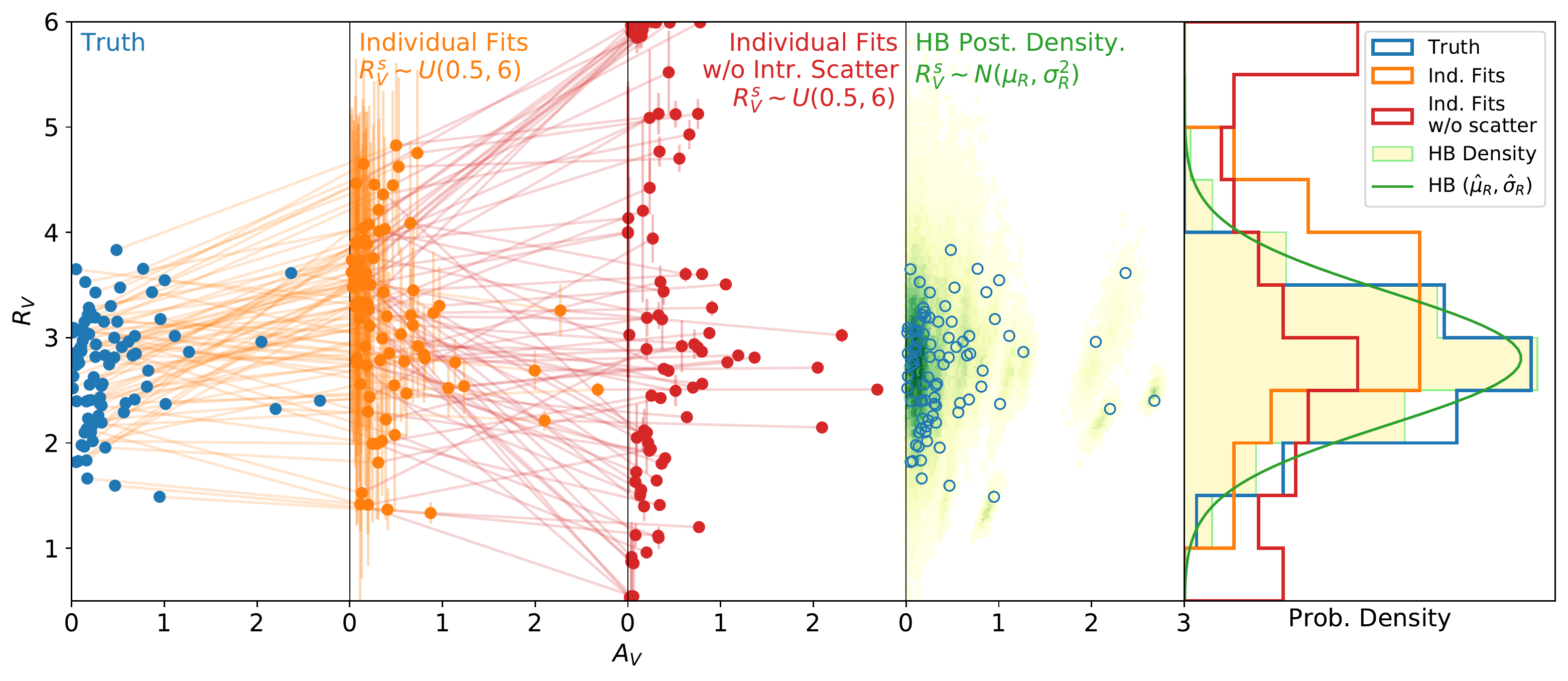}
    \caption{Estimation of $R_V$ and its population distribution for a simulated set of 86 SNe Ia. Blue points show the true simulated $R_V^s$ values for the individuals in the sample, with population mean $\mu_R=2.7$ and standard deviation $\sigma_R=0.5$. Orange points show $R_V^s$ estimates obtained from fitting each SN independently whilst assuming a uniform prior $R_V^s\sim U(0.5,6.0)$. These fits were carried out with \textsc{BayeSN}'s residual scatter model turned on. Red points show $R_V^s$ estimates obtained from fitting each SN independently with \textsc{BayeSN}'s residual scatter model turned off. Points are sorted by extinction, $A_V^s$, and estimates for a given individual are connected by faint lines. The yellow--green hexagonally-binned density plot shows the combined density of all individuals' $(A_V^s, R_V^s)$ posterior samples from the full hierarchical Bayesian analysis. The true $R_V^s$ values (same as in the leftmost panel) are overlaid as open blue circles. The rightmost panel shows several representations of the population distribution: a histogram of the sample's true $R_V^s$ values (blue line), a histogram of the individual point estimates (orange line), a histogram of the individual point estimates made without considering intrinsic scatter (red line), the marginal density of the individual $R_V^s$ posterior samples (yellow/green histogram), and the Gaussian PDF implied by the posterior mean estimate of the population hyperparameters ($\hat{\mu}_R$, $\hat{\sigma}_R$) from the hierarchical Bayesian analysis (green curve). From this, it is clear that the hierarchical Bayesian approach does the best job of capturing the true population distribution, whereas the point estimates from the individual fits are overdispersed, especially when intrinsic scatter is neglected in the fit.}
    \label{fig:lh_rv}
\end{figure*}

Figure \ref{fig:lh_rv} compares the individual $R_V^s$ estimates from the individual fits and the full hierarchical Bayesian analysis to the true input values of the simulated population. This comparison is made for the simulation with true $\mu_R=2.7$ and $\sigma_R=0.5$. We can see from this figure that the $R_V^s$ point estimates from the individual light curve fits (orange points in the second panel) are visibly overdispersed in comparison to the true population distribution (blue points in the leftmost panel), even when \textsc{BayeSN}'s residual scatter model is turned on. For low $A_V^s$, the $R_V^s$ estimates are more weakly constrained, and are biased slightly high (towards the prior mean). The individual uncertainties are large, though, except for the more highly extinguished individuals where the recovered $R_V^s$ estimates deviate less from the true values. When the fits are carried out with the residual intrinsic colour scatter turned off (red points in the third panel), the level of overdispersion is much greater, and the uncertainties on the individuals' $R_V^s$ and $A_V^s$ point estimates are much smaller in most cases.

In the fully hierarchical Bayesian analysis \ref{itm:full}, we obtain a set of posterior draws where the $i$th MCMC sample corresponds to a set of population hyperparameters $(\mu_R^i, \sigma_R^i)$, and an associated realisation, $\{R_V^s\}_i$, of individual $R_V^s$ values. In the fourth panel of Figure \ref{fig:lh_rv}, we show a hexagonally-binned probability density plot derived from the posterior samples of every individual's $(A_V^s, R_V^s)$, with overlaid blue circles showing the truth. At high extinction, the clusters of posterior samples corresponding to individual supernovae can be distinguished, whilst at lower extinction the samples blend together into a continuum that visually traces the true $(A_V, R_V)$ distribution of the sample. The rightmost panel of the plot shows the distribution of true $R_V^s$ (blue histogram), compared to the distribution of individual point estimates obtained with and without \textsc{BayeSN}'s intrinsic scatter model (orange and red histograms, respectively). The orange histogram is clearly broader than the blue, and is biased towards the prior mean of 3.25. The red is very overdispersed, with spurious overdensity near the prior bounds. The pale yellow/green histogram is a marginalisation of the density plot from one panel to the left -- i.e.\ the marginal density of all individuals' $R_V^s$ posterior samples obtained from the hierarchical Bayesian analysis. This traces the true $R_V$ distribution much better than the distribution of point estimates obtained from the supernova-by-supernova fits. Also shown in the rightmost panel is the Gaussian population distribution implied by the posterior mean estimate of the population hyperparameters $(\hat{\mu}_R=2.80, \hat{\sigma}_R=0.63)$ obtained from the hierarchical Bayesian analysis. As expected, this traces the true population distribution very well, being very close to the input values of the population hyperparameters $(\mu_R=2.7, \sigma_R=0.5)$. The hierarchical Bayesian analysis provides the best estimate of the true underlying $R_V$ population distribution.

\section{Results}
\label{sec:results}
In this section, we present the results of our analysis applied to the sample selected as described in \S\ref{sec:photodata}. In \S\ref{sec:step}, we describe results for the Hubble residuals and mass step for the SNe~Ia within the apparent $B-V\leq0.3$ colour cut. This colour cut is consistent with typical cosmological analyses, and is applied using apparent colour estimates from fits to the light curves. Eleven SNe are removed under this cut, yielding a low-to-moderate reddening sample of 75. Conclusions about the mass step are not strongly sensitive to this colour cut. In \S\ref{sec:dustlaws}, we investigate the inferred distribution of host galaxy dust laws over the full apparent colour range. For the full sample, NIR data near maximum are not required. In \S\ref{sec:nirmax}, results for a ``gold-standard'' subset with NIR near maximum light are described. In \S\ref{sec:highreddening}, we discuss the most highly-reddened individual objects in the sample.

\subsection{Hubble Residual Step}
\label{sec:step}
\subsubsection{Using NIR Data}
\label{sec:nirstep}
Our NIR Hubble residuals, obtained using the fixed $R_V$ fitting configuration, have a total root mean square (RMS) scatter of 0.137~mag (without correcting for any possible mass step), for the subset with apparent $B-V\leq0.3$. Estimated mass steps are included in Table \ref{tab:mass_steps_w_c_cut}. Correcting for the expected dispersion that would be contributed by peculiar velocity uncertainty (assuming $\sigma_\text{pec}=150$~km\,s$^{-1}$), we estimated a Hubble diagram scatter of $\sigma_{-\text{pv}}=0.121$~mag (using equation 33 of \citealp{mandel20}). Note that for this sample of 75 SNe Ia with apparent $B-V\leq0.3$, NIR data near maximum are not required. Results for a ``gold-standard'' subset with NIR near maximum light are described in Section  \ref{sec:nirmax}.

At a host galaxy stellar mass of $10^{10}~\mathrm{M}_\odot$, we estimate a Hubble residual step of $0.020\pm0.052$~mag, consistent with zero. At the median mass of the full sample ($10^{10.57}~\mathrm{M}_\odot$), we estimate a similar Hubble residual step of $0.050\pm0.032$~mag, consistent both with zero as well as a larger value, as seen by, e.g., \citet{uddin20}. Using the maximum likelihood estimator detailed in Appendix A of \citet{thorp21}, we estimate a preferred step location of $10^{10.77}~\mathrm{M}_\odot$. This is towards the higher end of reported estimates in the literature (although the sample being studied here favours high-mass hosts in general), and is similar to the step locations reported by, e.g., \citet{ponder20} and \citet{kelly10}. At this step location, our estimated NIR Hubble residual step size is $0.084\pm0.035$~mag, somewhat larger than the step at the median host mass.

\begin{table*}
    \centering
    \begin{threeparttable}
        \caption{Hubble residual magnitude steps for different fitting configurations and step locations, computed for the 75 SNe~Ia within the $B-V\leq0.3$ colour cut\tnote{a}. The dust parameters are determined from colours alone, without external distance information.}
        \label{tab:mass_steps_w_c_cut}
        \begin{tabular}{l l l c c c c c c}\toprule
             Config.\tnote{b}\ & $R_V$ treatment & Passbands & \multicolumn{3}{c}{Step size \tnote{c}} & RMS\tnote{d} & $\sigma_{-\text{pv}}$\tnote{e} & RMS$_{z>0.01}$\tnote{f}\\
             \cmidrule(lr){4-6}
             & & & $10^{10}~\mathrm{M}_\odot$ & $10^{10.57}~\mathrm{M}_\odot$ & $10^{10.77}~\mathrm{M}_\odot$\tnote{g} &\\\midrule
             \ref{itm:m20} & \emph{Fixed} ($R_V=2.89$) & $BVriYJH$ & $0.071\pm0.037$ & $0.062\pm0.026$ & $0.092\pm0.028$ & 0.123 & 0.108 & 0.113\\
             & & $BVri$ & $0.080\pm0.041$ & $0.078\pm0.028$ & $0.112\pm0.030$ & 0.150 & 0.137 & 0.141\\
             & & $YJH$ & $0.020\pm0.052$ & $0.050\pm0.032$ & $0.084\pm0.035$ & 0.137 & 0.121 & 0.122\\
             \ref{itm:pp} & \emph{Population\ $R_V$} & $BVriYJH$ & $0.064\pm0.038$ & $0.064\pm0.026$ & $0.090\pm0.028$ & 0.118 & 0.101 & 0.107\\
             \ref{itm:binned} & \emph{Colour-binned pop.\ $R_V$} (2-bin) & $BVriYJH$ & $0.062\pm0.038$ & $0.058\pm0.026$ & $0.084\pm0.029$ & 0.119 & 0.103 & 0.109\\
             & \emph{Mass-binned pop.\ $R_V$} ($10^{10}~\mathrm{M}_\odot$) & $BVriYJH$ & $0.062\pm0.038$ & $0.062\pm0.026$ & $0.088\pm0.028$ & 0.118 & 0.101 & 0.107\\
             & \emph{Mass-binned pop.\ $R_V$} ($10^{10.57}~\mathrm{M}_\odot$) & $BVriYJH$ & $0.059\pm0.038$ & $0.056\pm0.026$ & $0.082\pm0.028$ & 0.117 & 0.101 & 0.107\\
             \ref{itm:bs21} & \emph{Free, \citetalias{brout20} prior} & $BVriYJH$ & $0.045\pm0.038$ & $0.056\pm0.027$ & $0.081\pm0.029$ & 0.121 & 0.105 & 0.111\\
             & & $BVri$ & $0.030\pm0.044$ & $0.063\pm0.031$ & $0.096\pm0.033$ & 0.157 & 0.139 & 0.143\\
             \ref{itm:u} & \emph{Free, uniform prior} & $BVriYJH$ & $0.056\pm0.039$ & $0.055\pm0.027$ & $0.082\pm0.029$ & 0.127 & 0.113 & 0.118\\
             & & $BVri$ & $0.069\pm0.046$ & $0.074\pm0.032$ & $0.108\pm0.035$ & 0.182 & 0.169 & 0.173\\
             \bottomrule
        \end{tabular}
        \begin{tablenotes}
            \item [a] RMS, $\sigma_{-\text{pv}}$, and step sizes are computed from the subsample of 75 SNe with apparent $B-V\leq0.3$. The population $R_V$ model configurations (\ref{itm:pp} and \ref{itm:binned}) are fit to the full sample of 86 SNe, but only the Hubble residuals of the $B-V\leq0.3$ are used in evaluating the mass step.
            \item [b] Fitting configuration (as listed in Section \ref{sec:fittingconfigs}).
            \item [c] Estimated Hubble residual step size at different host galaxy stellar masses.
            \item [d] Root mean square (RMS) scatter of the Hubble residuals (without correction for a mass step).
            \item [e] Hubble diagram scatter with expected contribution from peculiar velocity uncertainty ($\sigma_\text{pec}=150$~km\,s$^{-1}$) removed \citep[eq.\ 33]{mandel20}.
            \item [f] RMS of Hubble residuals of the part of the sample (67 SNe) with $z_\text{CMB}>0.01$. Presented as an alternative to $\sigma_{-\text{pv}}$ for mitigating the effect of peculiar velocity uncertainty. Conclusions about the Hubble residual step size do not change substantially under this cut.
            \item [g] Step size computed at the MLE step location (estimated to be $10^{10.77}~\mathrm{M}_\odot$ for all $BVriYJH$ configurations).
        \end{tablenotes}
    \end{threeparttable}
\end{table*}

\subsubsection{Using Optical Data}
\label{sec:optstep}
When fitting the samples' optical ($BVri$) light curves under the fixed $R_V$ configuration, the Hubble diagram RMS ($\sigma_{-\text{pv}}$) is 0.150~(0.137)~mag, within the colour cut, somewhat larger than for the equivalent NIR fits. At $10^{10}~\mathrm{M}_\odot$, our estimate of the Hubble residual step is $0.080\pm0.041$~mag, larger than our equivalent estimate of the step size in the NIR Hubble residuals. At the sample median mass ($10^{10.57}~\mathrm{M}_\odot$) and maximum likelihood step location ($10^{10.77}~\mathrm{M}_\odot$), we estimate step sizes of $0.078\pm0.028$~mag and $0.112\pm0.030$~mag, respectively. A representative set of results are included in Table \ref{tab:mass_steps_w_c_cut}.

Under the free $R_V$ fitting configurations, we allow $R_V^s$ to be a free parameter in each fit (with one of two choices of fixed prior, and no pooling across the sample -- see Section \ref{sec:fittingconfigs}~\ref{itm:bs21} and \ref{itm:u} for details), and marginalise over it when estimating distances. For both choices of prior, the Hubble diagram RMS is $\sim0.16$--0.18~mag under a $B-V\leq0.3$ colour cut -- larger than the equivalent results under the configuration where a fixed $R_V=2.89$ was used. Our estimated Hubble residual magnitude steps for both prior choices are broadly consistent with one another. The estimated step sizes ($\gtrsim0.06$~mag at the median, $\gtrsim0.1$~mag at the MLE) are consistent with estimates made using the fixed $R_V$ fitting configuration. Detailed results are listed in Table \ref{tab:mass_steps_w_c_cut}. The two choices of $R_V$ prior we adopt are both only weakly constraining -- even though the \citetalias{brout20} population distributions (Eq.\ \ref{eq:bs21rvprior}) are more informative than the $U(1,6)$ prior, they still permit a wide range of $R_V$ for both mass bins due to their large $\sigma_R$ values. The combination of broad priors, and the limited $R_V$ constraining power of the optical data alone, contribute to more uncertain distance estimates, and the large Hubble residual scatter that we see here. It is noticeable that when optical and NIR data are combined (c.f.\ Section \ref{sec:optnirstep}), the free $R_V$ fitting configurations are competitive with the others.

\subsubsection{Using Optical + NIR Data}
\label{sec:optnirstep}
When estimating Hubble residuals from the full optical + NIR light curves, we adopt a variety of fitting configurations differing mainly in their treatment of $R_V$. Using the default \citetalias{mandel20} \textsc{BayeSN} model with a fixed $R_V=2.89$, we estimate a Hubble diagram RMS ($\sigma_{-\text{pv}}$) of 0.123~(0.108)~mag within the $B-V\leq0.3$ colour cut. At a host galaxy mass of $10^{10}~\mathrm{M}_\odot$, we estimate a Hubble residual step of $0.071\pm0.037$~mag. At the median host mass ($10^{10.57}~\mathrm{M}_\odot$) and MLE step location ($10^{10.77}~\mathrm{M}_\odot$), we find Hubble residual steps of $0.062\pm0.026$ and $0.092\pm0.028$~mag, respectively. These results are summarised in the first row of Table \ref{tab:mass_steps_w_c_cut}.

\begin{figure}
    \centering
    \includegraphics[width=\linewidth]{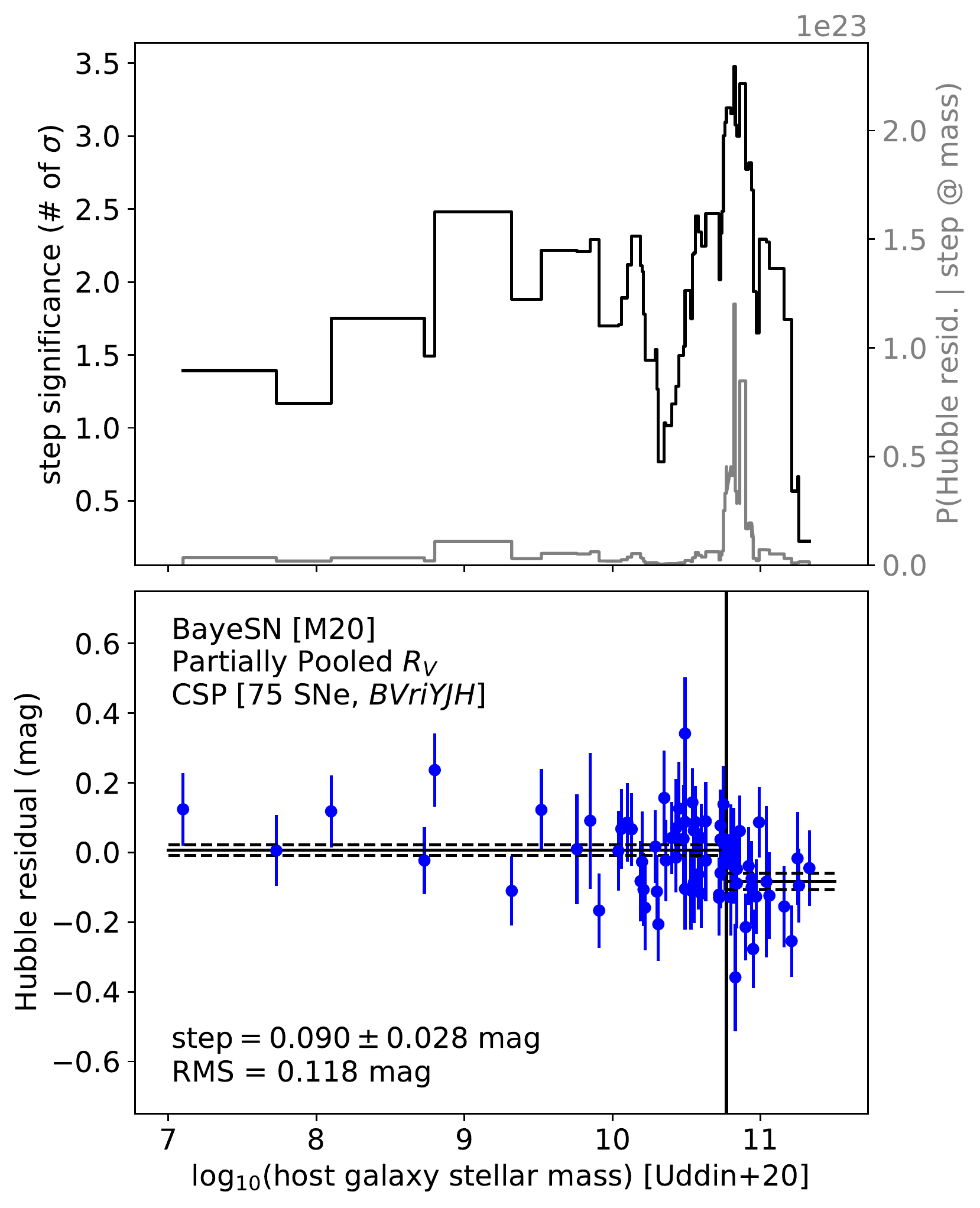}
    \caption{Mass step in Hubble residuals derived from photometric distance fits to optical+NIR CSP data under the population $R_V$ configuration with partial pooling, as per Section \ref{sec:fittingconfigs}~\ref{itm:pp}. (lower panel) Hubble residuals are shown in blue. Error bars are from quadrature sum of photometric distance uncertainty and uncertainty on external distance estimate (see Section \ref{sec:resids}). Mass step shown is at maximum likelihood step location. (upper panel) Black line and left hand axis show Hubble residual step significance vs.\ host mass. Grey line and right hand axis show marginal likelihood of Hubble residual step as a function of host mass.}
    \label{fig:step}
\end{figure}

For the population $R_V$ fitting configurations, the individual $R_V$ values are partially pooled: either across the full sample, or subsamples binned by host galaxy mass or SN Ia apparent colour (see Section \ref{sec:fittingconfigs} for full details). Across all different assumptions about the $R_V$ population distribution, the results are highly consistent. The Hubble residual RMS ($\sigma_{-\text{pv}}$) is typically $\sim0.12$~(0.10)~mag. At $10^{10}~\mathrm{M}_\odot$ we estimate a Hubble residual step of $\sim0.06$~mag. At the sample median host mass, we estimate Hubble residual steps of $\sim0.06$~mag (2.2--$2.5\sigma$) depending on the exact configuration considered. The most significant ($0.064\pm0.026$) occurs when the $R_V$ population is modelled as a Gaussian distribution for the full sample, and its hyperparameters are inferred. The least ($0.056\pm0.026$) corresponds to the case where the sample was divided equally into two bins of host galaxy stellar mass, and the $R_V$ population distribution in each is modelled as an independent Gaussian distribution, whose hyperparameters are inferred. For all population $R_V$ configurations, the preferred step location is at $10^{10.77}~\mathrm{M}_\odot$, where the step size is consistently $\gtrsim0.08$~mag ($\gtrsim2.9\sigma$). The lower panel of Figure \ref{fig:step} shows (within an apparent $B-V\leq0.3$ colour cut) the Hubble residuals and estimated mass step resulting from the unbinned population $R_V$ fitting configuration. The upper panel shows the estimated step significance at different masses (\`a la \citealp{roman18}, fig.\ 19; \citealp{kelsey21}, fig.\ 8), along with the marginal likelihood of the Hubble residuals conditional on a step at a particular mass.

In free $R_V$ fitting configurations, we fit every supernova independently, with one of two choices of prior -- either uniform, or a mass dependent prior motivated by \citetalias{brout20}. For both choices of prior, the results are very similar to the population $R_V$ configuration results, where $R_V$ was partially pooled. The Hubble diagram RMS is around 0.12--0.13~mag for both choices of prior, under the $B-V\leq0.3$ cut. At the median host galaxy mass, we estimate a Hubble residual step of $\sim0.06\pm0.03$~mag. At the MLE mas step location, our estimated Hubble residual steps are $\sim0.08\pm0.03$~mag. The Hubble diagram scatter is very close to the scatter achieved with all other optical + NIR configurations.

Results for all fitting configurations and step locations are presented in Table \ref{tab:mass_steps_w_c_cut}.

\subsection{Distribution of Dust Laws: Using Optical + NIR Data}
\label{sec:dustlaws}
\subsubsection{Constraints from Optical-NIR Colours: Fitting without External Distance Information}
\label{sec:optnirnoextern}
From the population $R_V$ fitting configurations with partial pooling, we can make inferences about the $R_V$ population distribution in the CSP sample. We marginalise over all distances and other supernova-level parameters (including the $R_V$ values for the constituent individuals of the sample) to estimate the posterior distribution of the hyperparameters defining the $R_V$ population distribution(s). Table \ref{tab:pop_RV_params_optnir} lists posterior estimates of these hyperparameters ($\mu_R$ and $\sigma_R$; population mean and standard deviation, respectively) for the model configurations where a truncated normal $R_V$ population distribution is assumed across either the full sample, or subsamples binned by colour or host galaxy mass. Our results for non-Gaussian population distributions are discussed in Appendix \ref{app:altdistributions}.

\begin{table*}
    \centering
    \begin{threeparttable}
        \caption{Inferred dust $R_V$ population distribution parameters from $BVriYJH$ data.}
        \label{tab:pop_RV_params_optnir}
        \begin{tabular}{l c c c c c c}\toprule
            & & & \multicolumn{2}{c}{Without distances\tnote{a}} & \multicolumn{2}{c}{With distances\tnote{b}}\\
            \cmidrule(lr){4-5} \cmidrule(l){6-7}
            Binning\tnote{c} & Subsample\tnote{d} & $N$\tnote{e} & $\mu_R$\tnote{f} & $\sigma_R$\tnote{g} & $\mu_R$ & $\sigma_R$\\\midrule
            Unbinned & - & 86 & $2.48\pm0.08$ & $0.34\pm0.11$ & $2.43\pm0.07$ & $0.31\pm0.10$\\
            Colour (2-bin) & $B-V\leq0.3$ & 75 & $2.59\pm0.14$ & $0.62\pm0.16$ & $2.50\pm0.12$ & $0.47\pm0.14$\\
            & $B-V>0.3$ & 11 & $2.39\pm0.08$ & $0.17\pm0.10$ & $2.34\pm0.08$ & $0.18\pm0.11$\\
            Colour (3-bin) & $B-V\leq0.3$ & 75 & $2.59\pm0.14$ & $0.57\pm0.20$ & $2.49\pm0.11$ & $0.47\pm0.14$\\
            & $0.3<B-V\leq0.7$ & 7 & $2.38\pm0.22$ & 0.34 (1.02) & $2.44\pm0.17$ & 0.40 (0.77)\\
            & $B-V>0.7$ & 4 & $2.36\pm0.19$ & 0.31 (0.76) & $2.29\pm0.16$ & 0.23 (0.75)\\
            Mass ($10^{10}~M_\odot$) & low & 15 & $2.56\pm0.22$ & $0.46\pm0.26$ & $2.59\pm0.22$ & $0.50\pm0.27$\\
            & high & 71 & $2.48\pm0.11$ & $0.37\pm0.15$ & $2.37\pm0.08$ & $0.28\pm0.11$\\
            & low; $B-V\leq0.3$\tnote{h} & 10 & $2.78\pm0.70$ & 1.62 (3.01) & $2.75\pm0.57$ & 1.18 (2.66)\\
            & high; $B-V\leq0.3$ & 65 & $2.54\pm0.15$ & $0.56\pm0.18$ & $2.46\pm0.13$ & $0.47\pm0.14$\\
            Mass ($10^{10.57}~M_\odot$) & low & 43 & $2.60\pm0.12$ & $0.33\pm0.12$ & $2.57\pm0.11$ & $0.31\pm0.11$\\
            & high & 43 & $2.35\pm0.14$ & $0.41\pm0.20$ & $2.26\pm0.11$ & $0.34\pm0.18$\\
            & low; $B-V\leq0.3$ & 36 & $2.79\pm0.18$ & $0.42\pm0.24$ & $2.75\pm0.15$ & $0.40\pm0.21$\\
            & high; $B-V\leq0.3$ & 39 & $2.35\pm0.27$ & $0.74\pm0.36$ & $2.26\pm0.19$ & $0.58\pm0.28$\\
            \bottomrule
        \end{tabular}
        \begin{tablenotes}
            \item [a] Constraints on $R_V$ distribution parameters from optical and NIR data without using external distance information.
            \item[b] Constraints with external distance information applied.
            \item [c] Binning strategy adopted (see Section \ref{sec:fittingconfigs}~\ref{itm:binned}). When binned by host galaxy stellar mass, the value in parentheses is the chosen split point. ``Unbinned'' corresponds to the population $R_V$ fitting configuration described in Section \ref{sec:fittingconfigs}~\ref{itm:pp}.
            \item [d] Subsample to which the parameters in columns 4 and 5 pertain. When binned by mass, ``low'' and ``high'' denote, respectively, the subsamples above and below the split point indicated in column 1.
            \item [e] Number of supernovae in each subsample.
            \item [f] Values quoted are posterior mean and standard deviation.
            \item [g] Values quoted are either posterior mean $\pm$ standard deviation, or 68 (95)\% upper bounds.
            \item [h] These rows correspond to the $B-V\leq0.3$ bin of a fitting configuration where mass and colour binning was applied.
        \end{tablenotes}
    \end{threeparttable}
\end{table*}

\begin{figure}
    \centering
    \includegraphics[width=\linewidth]{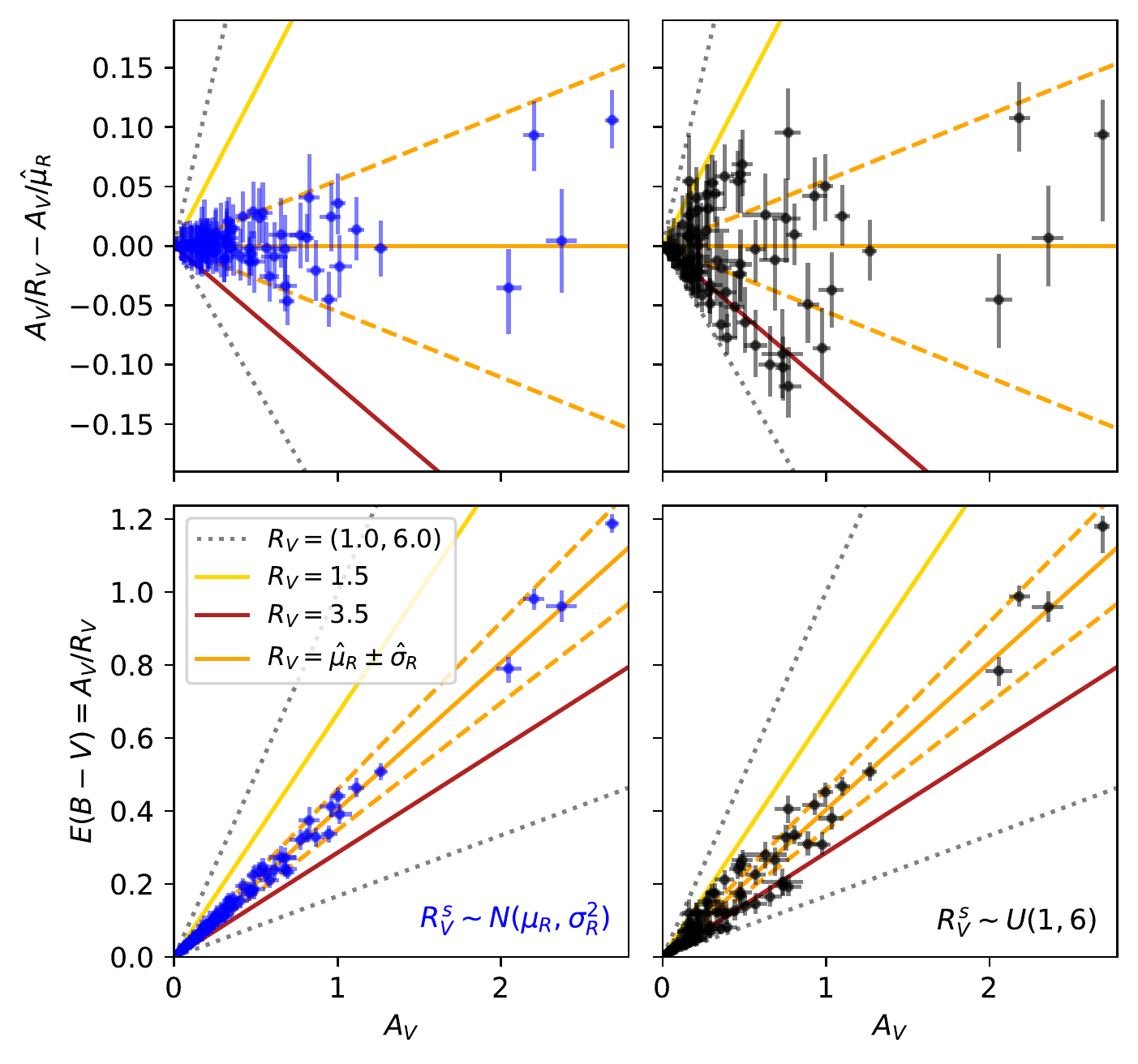}
    \caption{Constraints on the dust latent parameters $(A_V^s,R_V^s)$ for the 86 SNe in our sample. Blue markers show marginal posterior estimates (medians and 68 per cent credible intervals) from the population $R_V$ fitting configuration with an inferred Gaussian population distribution, $R_V^s\sim N(\mu_R,\sigma_R^2)$. Black markers show estimates from the free $R_V$ fitting configuration, with no pooling and independent uniform priors of $R_V^s\sim U(1,6)$. Orange solid and dashed lines show the $E(B-V)$ vs.\ $A_V$ relations ($\text{mean}\pm\text{std.\ dev}$) implied by the population distribution parameters $(\hat{\mu}_R=2.48$, $\hat{\sigma}_R=0.34)$ we estimate under partial pooling. (bottom panels) Constraints visualised in the space of derived $E_s(B-V)=A_V^s/R_V^s$ vs.\ $A_V^s$. (top panels) Residuals about the estimated population mean, i.e.\ the line $E(B-V)=A_V/\hat{\mu}_R=A_V/2.48$.}
    \label{fig:individuals}
\end{figure}

From the optical and NIR data of our full sample of 86 SNe Ia, we fit a truncated Gaussian population distribution for $R_V$ (Eq.\ \ref{eq:rvprior}) with mean $\mu_R=2.48\pm0.08$, and standard deviation $\sigma_R=0.34\pm0.11$. These parameters are inferred simultaneously with individual $R_V^s$ and $A_V^s$ values for the 86 SNe in our sample. The left hand panels of Figure \ref{fig:individuals} show our constraints on the $A_V^s$ and $R_V^s$ of the population's constituent individuals, in the space of derived reddening, $E^s(B-V)=A_V^s/R_V^s$, vs.\ extinction, $A_V^s$. These estimates were obtained under the population $R_V$ fitting configuration with partial pooling. The right hand panels contrast these with the constraints obtained under the free $R_V$ configuration with no pooling and a uniform prior, $R_V^s\sim U(1,6)$. In the latter case, the individual estimates show more apparent dispersion\footnote{Note that this does not translate into a significantly greater Hubble diagram dispersion (see Table \ref{tab:mass_steps_w_c_cut}), as the sensitivity of distance to misestimated $R_V$ is small when the dust extinction is low.} at low extinction, where the ability to constrain an individual's $R_V^s$ is poor. Hence, a significant portion of this apparent dispersion is likely due to individual estimation errors rather than true $R_V$ variation, as demonstrated in \S\ref{sec:why_hb}.

For the free $R_V$ configuration results, one could take the posterior means, $\hat{R}_V^s$, and standard deviations, $\hat{\sigma}_s$, as estimates of each supernova's $R_V^s$ (as in Section \ref{sec:why_hb}~\ref{itm:stddev}). If one then takes the standard deviation of the individuals' point estimates, one would naively estimate a population standard deviation of $\hat{\sigma}_R=0.81$, and a population mean of $\hat{\mu}_R=3.1$. This is similar to the approach taken by \citet{johansson21}, and yields a similar estimate of the population standard deviation (they estimate  $\hat{\sigma}_R=0.9$, on a sample which includes our own as a subset). An estimate of a population distribution width based on point estimates is likely to be an overestimate, however, as discussed in Section \ref{sec:why_hb} \citep[see also][]{loredo10, loredo19}. The optimal approach is to use a fully hierarchical model for the data. However, we can also apply the normal--normal model used in Section \ref{sec:why_hb}~\ref{itm:shrinkage} (Equation \ref{eq:8sch}) post-hoc to the $\{\hat{R}_V^s,\hat{\sigma}_s\}$ estimates from the free $R_V$ fits, to estimate the population parameters $(\mu_R,\sigma_R)$. From this, we estimate $\hat{\mu}_R=2.58$ and $\hat{\sigma}_R=0.31$, closer to the results from the fully hierarchical analysis under the population $R_V$ fitting configuration.

\begin{figure*}
    \centering
    \includegraphics[width=0.5\linewidth]{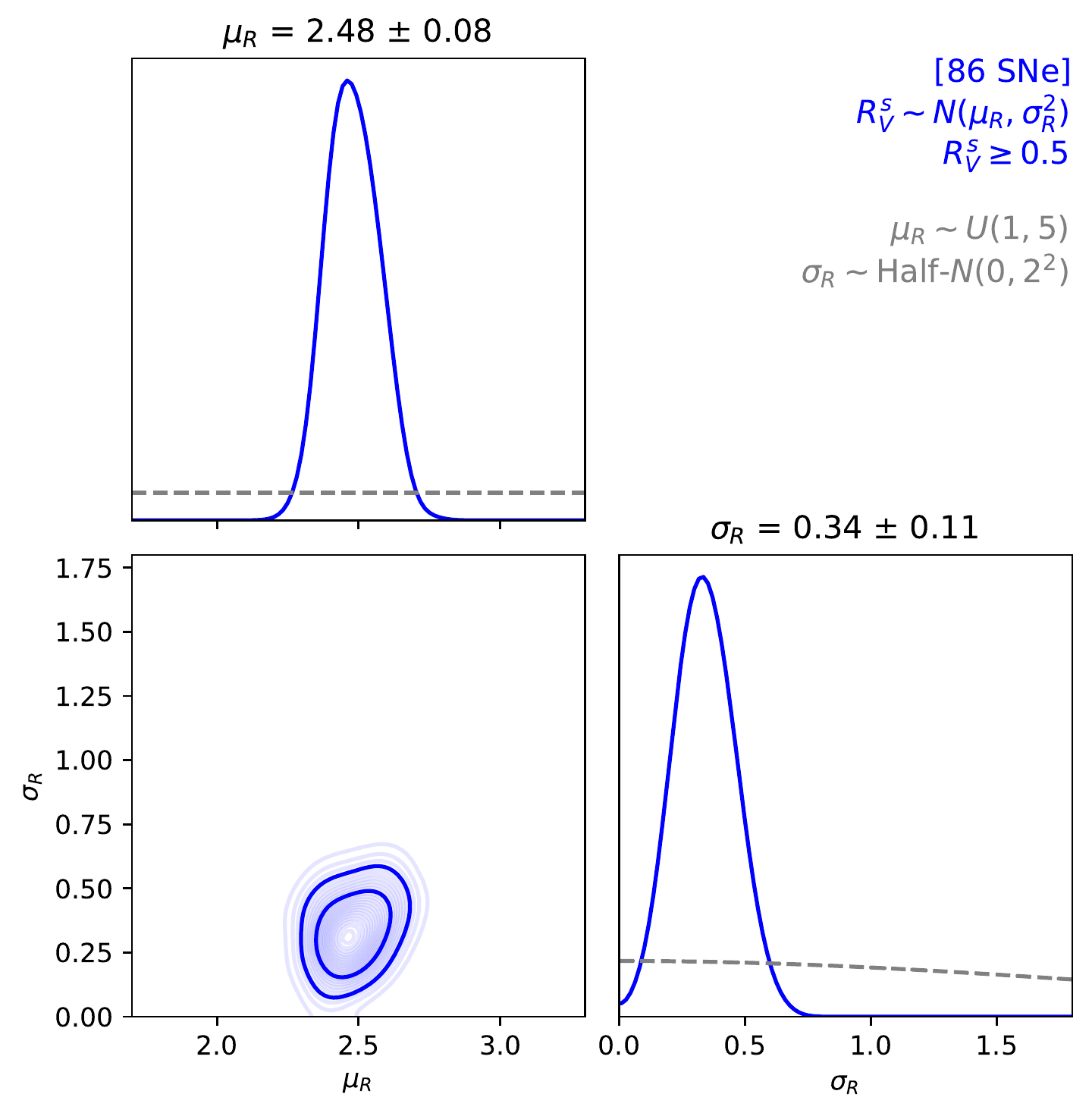}%
    \includegraphics[width=0.5\linewidth]{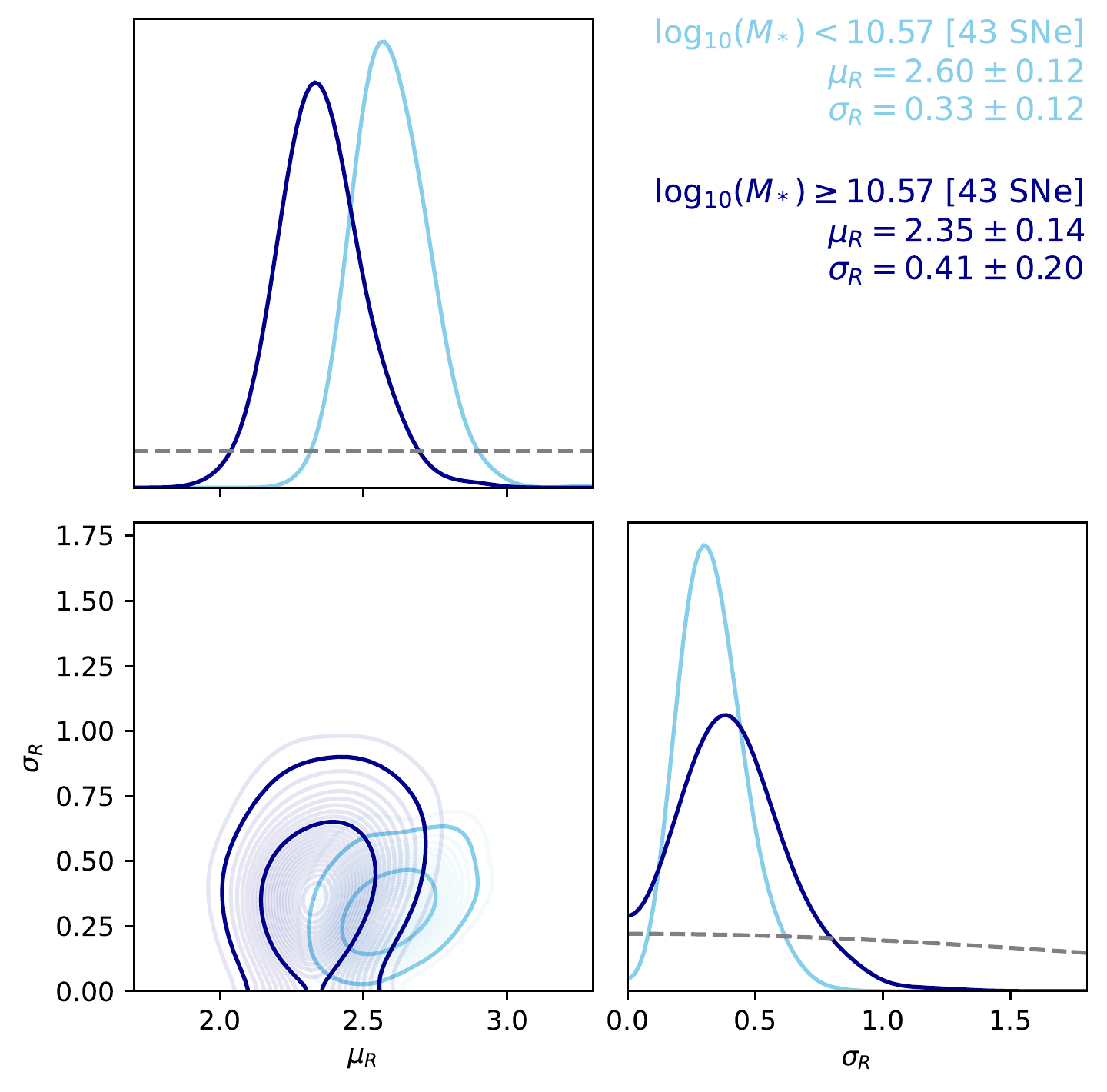}
    \includegraphics[width=0.5\linewidth]{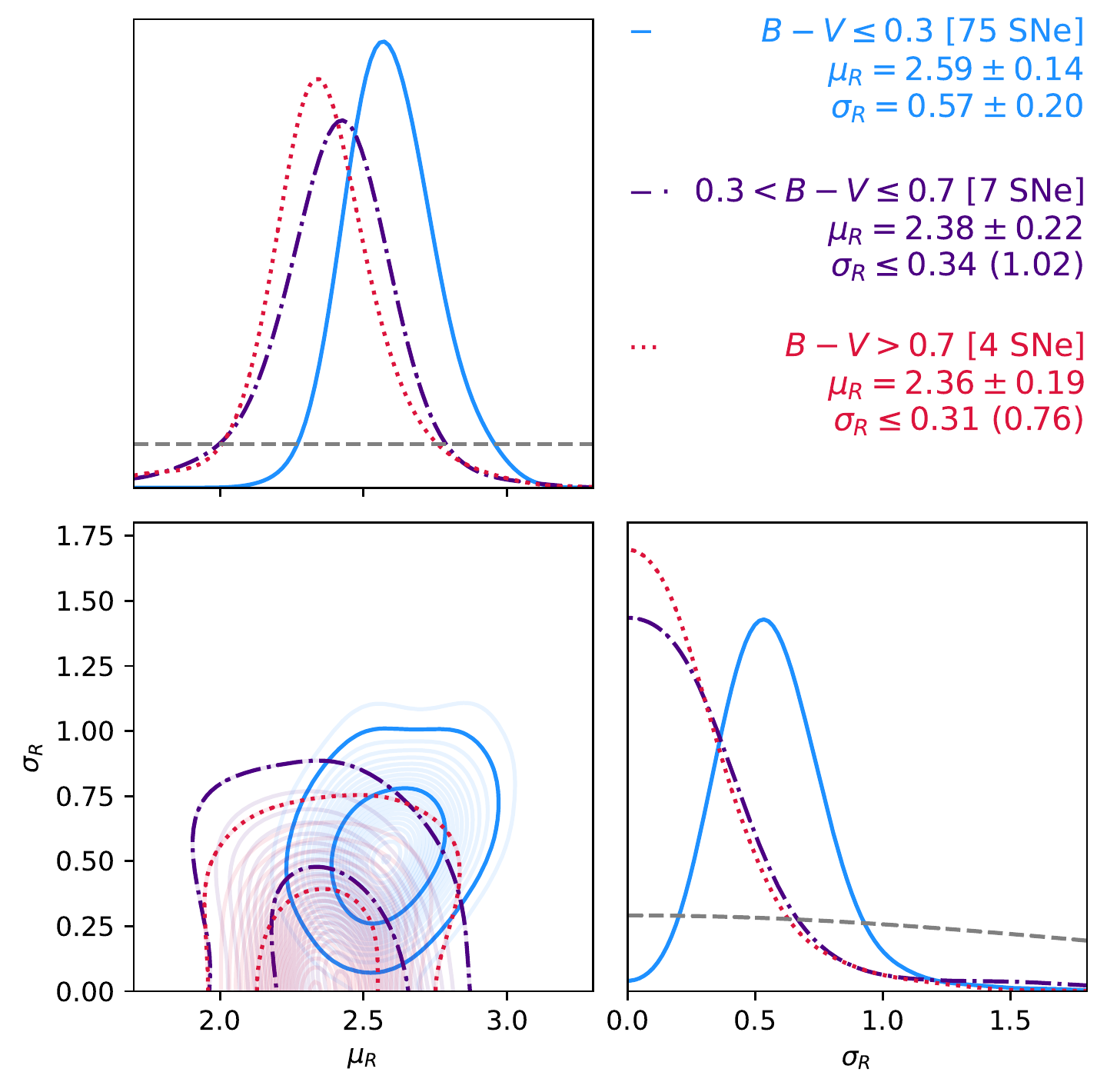}%
    \includegraphics[width=0.5\linewidth]{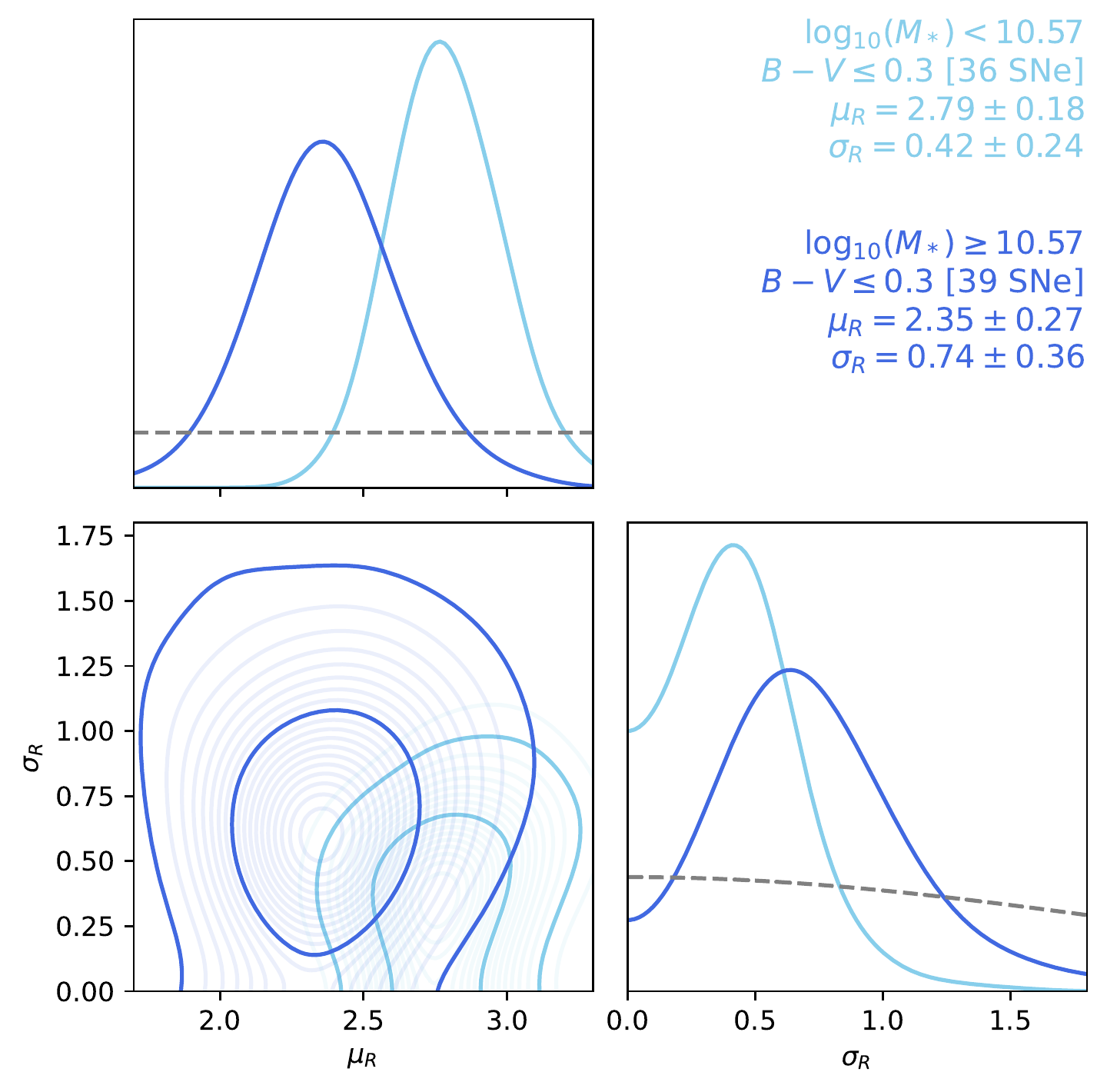}
    \caption{Two-dimensional marginal posteriors of $R_V$ population distribution parameters, $\mu_R$ and $
    \sigma_R$ (as defined in Eq.~\ref{eq:rvprior}), under partial pooling. Axis limits are the same for all quadrants. (upper left panels) Inferences from the population $R_V$ model with no binning by colour or mass, as per Section \ref{sec:fittingconfigs}~\ref{itm:pp}. (upper right panels) Inferences for the model with a population $R_V$ distribution within each of two separate bins of host galaxy stellar mass (above and below the sample median of $10^{10.57}~\mathrm{M}_\odot$), as per Section \ref{sec:fittingconfigs}~\ref{itm:binned}. (lower left panels) Same as upper right, but split into three bins of $B-V$ colour instead of two bins of host mass. (lower right panels) Same as upper right, but with the sample restricted to apparent $B-V\leq0.3$. }
    \label{fig:distributions}
\end{figure*}

The marginal posterior distribution of the population distribution parameters $(\mu_R,\sigma_R)$, obtained using the unbinned population $R_V$ fitting configuration, is shown in the upper left panel of Figure \ref{fig:distributions}. Our population hyperparameter estimates are consistent with the estimates made by \citet{thorp21} for SNe Ia from Foundation DR1 \citep{foley18,jones19}. As in \citet{thorp21}, our inferences show preference for a fairly small dispersion in $R_V$ ($\sigma_R<0.52$ with 95 per cent posterior probability). However, for the CSP sample considered here, we find much less posterior probability at $\sigma_R=0$ -- i.e.\ a preference for small but non-zero dispersion in $R_V$.

When the sample is split into either two or three bins by apparent $B-V$ colour, our estimates of the $R_V$ population mean $\mu_R$ are consistent between all colour bins (and with the unbinned result). There is a preference in the posterior for slightly lower $\mu_R$ for the SNe in the redder ($B-V>0.3$) bins, as reported by e.g.\ \citet{mandel11} and \citet{burns14}, although the difference here is not statistically significant. However, our sample has relatively few (eleven) SNe Ia with $B-V>0.3$, so our estimates of the $R_V$ population distribution in the higher colour bin(s) should be interpreted with this in mind. The lower left hand panels of Figure \ref{fig:distributions} show the posterior distributions of $(\mu_R,\sigma_R)$ in different colour bins for 3-bin model. Summaries of our posterior estimates of the population parameters are reported in Table \ref{tab:pop_RV_params_optnir}.

The most important part of the colour-binning exercise is the behaviour of the low--moderate reddening bin $B-V\leq0.3$ where the bulk of the sample (75 SNe) falls. The reddest SNe (even if they are few in number) exert considerable ``pull'' on the inferred $R_V$ distribution for a sample of SNe Ia \citep[see, e.g.][]{folatelli10,mandel11}. This is due to their larger lever arm distance from the intrinsic colour locus, which gives them much greater sensitivity to small changes in $R_V$. Splitting the sample by colour isolates the less reddened ($B-V\leq0.3$) supernovae from the influence of their highly reddened counterparts. Because of this, the strategy of binning by colour has the greatest effect on the $R_V$ distribution amongst the least reddened supernovae. For $B-V\leq0.3$, we estimate an $R_V$ population distribution with population mean $\mu_R=2.59\pm0.14$ and standard deviation $\sigma_R=0.62\pm0.16$. The estimated population mean $\mu_R$, for $B-V\leq0.3$, is consistent with the value estimated for the full sample without colour binning. The estimated population standard deviations, $\sigma_R$, for $B-V\leq0.3$ and the unbinned sample are not inconsistent. For the colour-restricted subsample, however, there is greater posterior probability towards larger standard deviations, with the 95th percentile falling at $\sigma_R\approx0.9$ (c.f.\ the unbinned result, where we estimate that $\sigma_R\lesssim0.5$ with 95 per cent posterior probability). This is likely driven by the fact that the restriction to $B-V\leq0.3$ removes the ability of the highly reddened supernovae (which seem to show fairly small $R_V$ dispersion in this sample -- see Table \ref{tab:pop_RV_params_optnir}, or red/purple contours in lower left panel of Fig.\ \ref{fig:distributions}) to pull the entire population distribution tighter.

When the sample is divided by host galaxy stellar mass, we estimate consistent $R_V$ distributions between the two mass bins when splitting at either $10^{10}$ or $10^{10.57}~\mathrm{M}_\odot$. When the sample is divided at the latter (the median host galaxy mass), we estimate population means of $\mu_R=2.60\pm0.12$ for lower-mass hosts, and $\mu_R=2.35\pm0.14$ for the higher-mass hosts. We estimate similar population standard deviations ($0.33\pm0.12$ and $0.41\pm0.20$ for low- and high-mass hosts, respectively) in the two mass bins, also. The upper right panel of Figure \ref{fig:distributions} overlays the $(\mu_R,\sigma_R)$ posterior distributions for the two mass bins. When dividing the sample at $10^{10}~\mathrm{M}_\odot$, the consistency between the two bins is even stronger (see Table \ref{tab:pop_RV_params_optnir}), albeit with the caveat that there are only 15 SNe Ia in the sample with a host mass $<10^{10}~\mathrm{M}_\odot$. The preference for similar $R_V$ population distributions for low- and high-mass host galaxies aligns with the results of \citet{thorp21} for the Foundation DR1 \citep{foley18,jones19} SN Ia sample.

This result also holds when we repeat our analysis with both host mass and colour binning. In this mode, we focus on the $B-V\leq0.3$ bin, as it is most cosmologically relevant, and most comparable to previous analyses \citep[e.g.][]{brout20,thorp21}. Additionally, it is even more difficult to draw conclusions about the sparsely populated $B-V>0.3$ bin when this is further divided by host mass. Our $R_V$ distribution inferences are consistent when splitting the $B-V\leq0.3$ subsample at either $10^{10}~\mathrm{M}_\odot$ or $10^{10.57}~\mathrm{M}_\odot$, and the results are consistent with those obtained without the colour cut. Using the split at $10^{10.57}~\mathrm{M}_\odot$, we find a population mean $R_V$ of $\mu_R=2.79\pm0.18$ for low-mass hosts and $\mu_R=2.35\pm0.27$ for high-mass hosts, with population standard deviations of $\sigma_R=0.42\pm0.24$ and $\sigma_R=0.74\pm0.36$, respectively. The lower right panels of Figure \ref{fig:distributions} show the joint posterior distributions over $(\mu_R,\sigma_R)$ for the two mass bins in the case. Our estimates of the population means are consistent to within $1.4\sigma$, and are consistent with the results obtained by \citet{thorp21} on the completely independent Foundation dataset. Our complete results for the $10^{10}~\mathrm{M}_\odot$ and $10^{10.57}~\mathrm{M}_\odot$ mass splits, with $B-V\leq0.3$ colour cut can be found in Table \ref{tab:pop_RV_params_optnir}.

Although these analyses were carried out without any external constraints on the distance moduli of individual SNe, repeating our fits with external distance constraints imposed makes very little difference to the conclusions about the dust law population distribution(s). The next section \ref{sec:optnirextern} discusses this. Our $R_V$ distribution inferences using optical data only are presented in Appendix \ref{sec:optdustlaws}.

\subsubsection{Constraints from Optical--NIR Colours and Luminosity: Using External Distance Information}
\label{sec:optnirextern}
In Section \ref{sec:optnirnoextern}, the dust law population parameters are inferred during photometric distance estimation, in which the photometric distances are marginalised over without external distance constraints being imposed. Under that mode of analysis, information about $R_V$ comes from colours (i.e. the relative brightness differences between photometric passbands). In this section, we repeat our $R_V$ population distribution analysis, but with external distance constraints (see Section \ref{sec:resids}) imposed. When the distance moduli are strongly constrained externally, information about $R_V$ comes from from colours and absolute magnitudes, rather than colours alone\footnote{Figure 12 of \citet{mandel20} provides a helpful low-dimensional visualisation of how SN Ia colour--colour information can be used to infer $R_V$ \citep[see, also, e.g.][fig.\ 13]{folatelli10}. Similarly, figure 5 of \citet{thorp21} visualises how colour--magnitude information can be used to similar effect.}. As such, this approach is closer to that of \citet{mandel20} and \citet{thorp21}, where $R_V$ (or its population distribution) was constrained during training of the \textsc{BayeSN} model, with external distance constraints imposed.

Imposing external distance constraints makes very little difference to our $R_V$ population distribution inferences when conditioning on the full $BVriYJH$ light curves. The two rightmost columns of Table \ref{tab:pop_RV_params_optnir} list our estimates of the population distribution parameters $(\mu_R,\sigma_R)$ from this analysis (c.f.\ the adjacent columns, listing the results when no external distance constraints were used). The consistency of these results indicates that when optical and NIR data are available, the colour information alone is highly informative about the distribution of $R_V$.

\subsection{A ``Gold-Standard'' Subsample with NIR at Maximum}
\label{sec:nirmax}
In this section, we present a limited set of results on a ``gold-standard'' subset of the CSP sample -- namely the 28 SNe Ia that were identified by \citet{avelino19} as having NIR data around maximum light (``NIR@max''). This subset is of particular interest, as we expect it to have the highest quality NIR light curves. Moreover, NIR observations near maximum light are of particular value for the study of host galaxy dust, as the intrinsic optical$-$NIR colour dispersion is particularly small here (see Fig.~\ref{fig:colour_curves}, and discussion in Section \ref{sec:why_onir}). This makes it easiest to separate the effect of dust from the intrinsic colour variation.

We perform a joint fit to the optical and NIR light curves of these 28 SNe, under the population $R_V$ fitting configuration, where $R_V$ is partially pooled across the sample. From this gold-standard subsample, we estimate a population mean $R_V$ of $\mu_R=2.88\pm0.22$, and a population standard deviation of $\sigma_R=0.46\pm0.26$. This population mean estimate is highly consistent with the global $R_V$ estimate ($R_V=2.89\pm0.20$) made in \citet{mandel20}, whose training set included these 28 SNe Ia as a significant subset. It is also consistent with the population mean estimate ($\mu_R=2.70\pm0.25$) obtained by \citet{thorp21} on the completely independent Foundation supernova sample. The population distribution we infer for the gold-standard subset of 28 SNe is consistent with our inference for our full $B-V\leq0.3$ subsample of 75 SNe, where we estimate $\mu_R=2.59\pm0.14$ and $\sigma_R=0.62\pm0.16$.

\begin{table}
    \centering
    \begin{threeparttable}
        \caption{Hubble residual scatter for the ``gold-standard'' NIR@max subsample of 28 SNe Ia.}
        \label{tab:nirmax}
        \begin{tabular}{l l c c c}\toprule
             Config.\tnote{a}\ & Passbands & RMS & $\sigma_{-\text{pv}}$ & RMS$_{z>0.01}$\tnote{b}\\\midrule
             \emph{Fixed $R_V$}\tnote{c} & $BVriYJH$ & 0.087 & 0.075 & 0.089\\
             & $BVri$ & 0.113 & 0.099 & 0.115\\
             & $YJH$ & 0.091 & 0.074 & 0.084\\
             \emph{Pop.\ $R_V$}\tnote{d} & $BVriYJH$ & 0.087 & 0.074 & 0.087\\
             \bottomrule
        \end{tabular}
        \begin{tablenotes}
            \item [a] Fitting configuration (see \S\ref{sec:fittingconfigs}).
            \item [b] Computed for the 25 SNe with $z_\text{CMB}>0.01$.
            \item [c] Fixed $R_V=2.89$.
            \item [d] $R_V$ partially pooled across the 28 SNe Ia. In this configuration, we infer a population mean of $\mu_R=2.88\pm0.22$, and std.\ dev.\ of $\sigma_R=0.46\pm0.26$.
        \end{tablenotes}
    \end{threeparttable}
\end{table}

As one would expect, we estimate a very small Hubble diagram scatter on this subsample. Table \ref{tab:nirmax} summarizes these results. Using the full optical and NIR light curves, and a fixed $R_V=2.89$ we estimate Hubble residual RMS ($\sigma_{-\text{pv}}$) of 0.087~(0.075)~mag -- very similar to the Hubble diagram scatter found in \citet{mandel20}, which included these 28 SNe as a significant subset of their fiducial sample. The RMS is almost exactly the same when using the NIR light curves only, indicating the power of having NIR data at peak where SNe Ia are very close to standard candles \citep[see e.g.][]{avelino19}. Using only the optical light curves, the RMS ($\sigma_{-\text{pv}}$) is slightly higher (0.113~(0.099)~mag) but still small, reflecting the fact that this subset generally has the highest quality light curves in all passbands. The Hubble diagram scatter from population $R_V$ fits with partial pooling is almost identical to the fits where $R_V$ was fixed. This is unsurprising, since we estimate that this subsample is highly consistent with a population mean $\mu_R\approx2.9$, with a narrow estimated $R_V$ distribution. Moreover, since this sample has $B-V\leq0.3$ (and thus low--moderate dust reddening) by construction, the sensitivity of distance estimates to small deviations of $R_V$ from the population mean will be fairly limited -- especially when NIR data are available.

\subsection{Individual Highly Reddened Objects}
\label{sec:highreddening}
In this section, we investigate the four most highly reddened SNe Ia in the sample: 2005A, 2006X, 2006br, and 2009I. These supernovae constitute the reddest subsample (apparent $B-V>0.7$) in our colour-binned analyses. Given their extremely high dust extinction, we expect precise inference of $R_V$ for these SNe, independently from the dust population distribution or prior, and will thus focus on the results where each supernova was fitted separately (i.e.\ no pooling) with a uniform prior, $R_V\sim U(1,6)$. These results made use of the full $BVriYJH$ light curves. Table \ref{tab:individual_RVs} lists our inferred $A_V$ and $R_V$ values for the four highly reddened SNe Ia. Also listed are the intrinsic $B-V$ colours that we estimate at maximum light, and a derived estimate of colour excess due to reddening: $E(B-V)=A_V/R_V$.

\begin{table}
    \centering
    \begin{threeparttable}
        \caption{Dust inferences\tnote{a} for the four most highly reddened SNe Ia.}
        \label{tab:individual_RVs}
        \begin{tabular}{l c c c c}\toprule
             SN & $(B-V)_\text{int}$\tnote{b} & $A_V$\tnote{c} & $R_V$\tnote{d} & $E(B-V)$\tnote{e}\\\midrule
             2005A & $0.16\pm0.03$ & $2.18\pm0.06$ & $2.21\pm0.10$ & $0.99\pm0.03$\\
             2006X & $0.09\pm0.03$ & $2.67\pm0.05$ & $2.24\pm0.07$ & $1.19\pm0.03$\\
             2006br & $0.06\pm0.04$ & $2.36\pm0.10$ & $2.46\pm0.14$ & $0.96\pm0.04$\\
             2009I & $-0.01\pm0.04$ & $2.06\pm0.08$ & $2.63\pm0.18$ & $0.78\pm0.04$\\
             \bottomrule
        \end{tabular}
        \begin{tablenotes}
            \item [a] All values quoted are posterior mean $\pm$ standard deviation.
            \item [b] Estimated in the CSP $B$- and $V$-bands at the time of $B$-band maximum.
            \item [c] Inferred under an exponential prior: $A_V\sim\exponential(\tau_A=0.33)$.
            \item[d] Inferred under a uniform prior: $R_V\sim U(1,6)$.
            \item [e] Derived from $E(B-V)=A_V/R_V$.
        \end{tablenotes}
    \end{threeparttable}
\end{table}

We can compare our estimates from Table \ref{tab:individual_RVs} to other $A_V$, $R_V$, and $E(B-V)$ estimates for these supernovae, when these are available in the literature. \citet{burns14} carried out an extensive study of the CSP data available at the time, including three of the four highly-extinguished SNe Ia included here (2005A, 2006X, 2006br). When assuming the \citet{fitzpatrick99} reddening law and a uniform prior on $R_V$, they estimate $R_V=(1.9\pm0.1, 1.8\pm0.1, 2.5\pm0.2)$ and $E(B-V)=(1.129\pm0.029, 1.360\pm0.026, 0.896\pm0.050)$ for supernovae (2005A, 2006X, 2006br). For 2006br, our estimates of $E(B-V)$ and $R_V$ agree very closely with those of \citet{burns14}. For 2005A, our estimate of $E(B-V)$ is lower, and our estimate of $R_V=2.21\pm0.10$ is somewhat higher than that of \citet{burns14}. For 2006X, our estimated $E(B-V)=1.17\pm0.05$ is significantly lower than the $E(B-V)=1.360\pm0.026$ estimated by \citet{burns14}, and our estimated $R_V=2.24\pm0.07$ is significantly higher than the \citet{burns14} estimate of $R_V=1.8\pm0.1$.

Elsewhere in the literature, \citet{folatelli10} also present estimates of $R_V$ based on optical and NIR CSP photometry for supernovae 2005A and 2006X, finding values of $1.68\pm0.10$ and $1.55\pm0.07$, respectively. Also using CSP photometry, \citet{phillips13} estimate $A_V=1.88^{+0.09}_{-0.13}$ and $R_V=1.31^{+0.08}_{-0.10}$ for 2006X. From their own optical and NIR photometry, taken on a variety of telescopes, \citet{wang08} estimate $R_V=1.48\pm0.06$ and $A_V=2.1$. By comparing the optical spectra of 2006X to the spectroscopically similar unreddened supernova 2004eo, \citet[ch.~5]{eliasrosa07} estimate $R_V=1.74\pm0.11$ and $A_V=2.08\pm0.57$ -- a slightly higher $R_V$ value than the photometric analyses of \citet{wang08}, \citet{folatelli10}, and \citet{phillips13}, but still lower than our own estimate. Spectropolarimetric observations of 2006X \citep[see e.g.][]{patat09,patat15} have also been used to constrain its line of sight $R_V$. \citet{patat15} report an upper limit of $R_V\lesssim2.2$, based on empirical linear relations \citep{serkowski75,whittet78,clayton88} between $R_V$ and the wavelength of maximum polarisation (estimated to be $\lambda_\text{max}\lesssim0.4~\upmu$m). Inserting the estimate of $\lambda_\text{max}=0.365\pm0.02~\upmu$m for 2006X \citep{patat15} directly into the $R_V$ vs.\ $\lambda_\text{max}$ relations from \citet{serkowski75}, \citet{whittet78}, and \citet{clayton88} gives $R_V=2.01\pm0.11$, $2.04\pm0.15$, and $2.15\pm0.86$, respectively. All of these would be consistent with our own estimate for 2006X ($R_V=2.24\pm0.07$).

In common between \citet{wang08}, \citet{folatelli10}, and \citet{phillips13} is their use of the \citet{cardelli89} dust law (or the update due to \citealp{odonnell94}), rather than the \citet{fitzpatrick99} law that we use here. The \citet{cardelli89} law typically yields lower $R_V$ estimates on a supernova-by-supernova basis than the \citet{fitzpatrick99} law, as discussed by \citet{burns14}\footnote{Indeed, for supernovae 2005A and 2006X, \citet{burns14} estimate $R_V$ values of $1.4\pm0.1$ and $1.3\pm0.1$, respectively. However, they also note that for these two supernovae, a CSM-motivated power-law in the style of \citet{goobar08} fits the data better than either of the \citet{cardelli89} or \citet{fitzpatrick99} dust laws. Whilst noting this caveat, we will defer a thorough investigation of dust-law functional form to future work.}. This could be the source of some of the discrepancy between our own $R_V$ estimate and these other works. A further cause for the discrepancy could be due to the assumptions about the intrinsic colours from which colour excess is determined. \citet{wang08}, \citet{folatelli10}, and \citet{phillips13} all primarily rely on an assumed or fitted linear relation between mean intrinsic colour and light curve decline rate ($\Delta m_{15}(B)$), \`a la \citet{phillips99}. As noted by \citet{wang08}, a significant deviation of a supernova's intrinsic colour from the population mean $(B-V)_\text{int}$ vs.\ $\Delta m_{15}(B)$ relation could compromise one's ability to correctly estimate the level of dust reddening. The \textsc{BayeSN} SED model we have used here ought to be guarded against such an issue, as it incorporates a residual scatter model, inferred during model training, that naturally allows for realistic variation about the mean intrinsic colours associated with a given light curve shape (see the black curves in Figure \ref{fig:colour_curves} for a visualisation of this).

It is noticeable that we estimate that 2005A and 2006X are both intrinsically very red, with $(B-V)_\text{int}>0.1$. This could be a possible cause for the discrepancy we see between our results and other photometric studies in the literature. In \citet{mandel20}, the inferred distribution of intrinsic $B-V$ colour at peak had a population mean of $-0.028\pm0.007$~mag, and population standard deviation of $0.065\pm0.005$~mag. This intrinsic colour distribution is implicitly embedded in the \citetalias{mandel20} \textsc{BayeSN} model (via the population distribution of the $\epsilon_s(t,\lambda)$ functions) we have used throughout the present work. Based on the intrinsic colour estimates in Table \ref{tab:individual_RVs}, 2005A and 2006X would both fall fairly far into the red tail of the \citetalias{mandel20} population distribution -- both beyond the $95$th percentile of said distribution. Our estimated intrinsic colours for 2006br and 2009I are less extreme -- both within the 95th percentile -- but still in the redder half of the \citetalias{mandel20} distribution.

Supernovae 2005A, 2006X, and 2006br have all been identified as having high ejecta velocities ($\gtrsim11800$~km\,s$^{-1}$), as measured by the \ion{Si}{ii} line at maximum light \citep{wang09, foley11, blondin12, folatelli13, dettman21}. Supernova 2009I, on the other hand, was identified by \citet{krisciunas17} as being ``normal'' (i.e.\ having \ion{Si}{ii} velocity $<11800$~km\,s$^{-1}$ at maximum light) based on the \citet{wang09} subclassification scheme. Reported \ion{Si}{ii} velocities for 2005A (based on the line at 6355~\AA) range between $13790\pm350$~km\,s$^{-1}$ \citep{dettman21}, and $14976\pm279$km\,s$^{-1}$ \citep{folatelli13}. For 2006X, \ion{Si}{ii} estimates are typically $\approx16000$~km\,s$^{-1}$ \citep{foley11, blondin12, folatelli13}, although \citet{dettman21} find a lower value of $13790\pm220$~km\,s$^{-1}$. For 2006br, \citet{foley11}, \citet{blondin12}, \citet{folatelli13}, and \citet{dettman21} all report ejecta velocity estimates of $\approx14000$~km\,s$^{-1}$. It has been reported by various past studies \citep[e.g.][]{foleykasen11, foley12, blondin12, mandel14} that SNe Ia with higher ejecta velocities may be intrinsically redder. Additionally, \citet{siebert20} report a Hubble residual vs.\ ejecta velocity step, that would be consistent with a velocity--colour relation. Such an effect could explain the very red colours estimated for 2005A, 2006X, and 2006br. All three of these SNe have estimated \ion{Si}{ii} velocities $\gtrsim14000$~km\,s$^{-1}$ -- this puts them towards the fastest tail of the velocity distributions that are typically observed (see e.g.\ fig.~1 in \citealp{foley11}). In a scenario where \ion{Si}{ii} velocity and intrinsic colour are correlated \citep[\`a la][]{foley11, foley12, mandel14}, one would thus expect these SNe to lie towards the redder tail of the intrinsic $B-V$ colour distribution, qualitatively consistent with our estimates in Table \ref{tab:individual_RVs}. Our inference that 2009I has an intrinsic colour closer to the population mean would also seem to align with this picture, as this SN is identified as having a normal \ion{Si}{ii} velocity \citep{krisciunas17}. A more quantitative investigation of this effect is beyond the scope of the present work, but could be an interesting topic for future investigation.

It is possible that these SNe are not as intrinsically red as our inferences suggest, so we consider the sensitivity of these results to the extinction prior. The results presented in Table \ref{tab:individual_RVs} were carried out using an exponential prior on extinction, $A_V\sim\exponential(\tau_A=0.33)$, inferred during the training of the \citetalias{mandel20} \textsc{BayeSN} model. A Bayesian calculation under such a prior will naturally penalise higher $A_V$ values, meaning it may be easier to make a supernova more intrinsically red to explain some of its apparent colour. To investigate whether these intrinsically red colours can be traced to the dust prior, we repeated our fits with an improper flat prior, $A_V\sim U(0,\infty)$, on positive dust extinction. We found that our inferences were essentially unchanged under this alternative prior. 

It is worth noting that, whilst we estimate $R_V\approx2.2$--2.7 for these four very highly reddened SNe Ia in the CSP-I sample, this may not be characteristic of all such SNe (see e.g.\ SN~2012cu with $E(B-V) \approx 1$, $R_V\approx2.8$--3.0; \citealp{amanullah15, huang17}).

\section{Conclusions}
\label{sec:conclusions}
In this paper, we have conducted a detailed investigation into the dust law $R_V$ distribution in SN Ia host galaxies, and tested the sensitivity of the mass step to a variety of Bayesian treatments of this distribution. As a precursor to our main analysis, we have argued for the importance of leveraging colour information across the full optical--NIR wavelength range in order to effectively separate the effects of dust from intrinsic SN Ia colour variation (\S\ref{sec:why_onir}). Moreover, we have presented a simulation-based demonstration of several of the statistical challenges inherent in inferring the host galaxy dust law $R_V$ distribution from a sample of SN Ia photometry (\S\ref{sec:why_hb}). In particular, we have shown that:
\begin{itemize}
    \item fitting SNe Ia one-by-one, first obtaining point estimates of $R_V$ for each individual with a uniform prior, and then inferring the population distribution by taking the simple standard deviation of the point estimates, will lead to a biased (high) inference of the $R_V$ dispersion within the population;
    \item if SN Ia residual intrinsic colour scatter is not accounted for in the fit, $R_V$ estimates for individual supernovae will be overconfident (individual uncertainties underestimated) and overdispersed (sample variance larger than the true population variance);
    \item a shrinkage estimator based on a simple hierarchical Bayesian approach can be used to obtain a more robust constraint on the $R_V$ distribution from a collection of point estimates, but \textit{only} if these have appropriate individual uncertainties (obtained by robustly accounting for intrinsic colour scatter); and
    \item a fully hierarchical Bayesian approach is the ideal solution, with the distribution of dust laws being constrained whilst fitting the full sample of light curves simultaneously.
\end{itemize}
We emphasize that these statistical effects will generally apply for any method of estimating dust properties from light curves or colour curves.  However, we demonstrate and address them here using the \textsc{BayeSN} SED model since it uniquely contains a quantitative description of the time-dependent optical--NIR residual intrinsic colour scatter that can be used both for simulation and fitting of SN Ia light curves with variable dust laws. Furthermore, \textsc{BayeSN} has the unique capability of jointly fitting ensembles of SN Ia light curves in a hierarchical configuration.

Motivated by these lessons, we have applied our \textsc{BayeSN} model for the SEDs of SNe Ia \citep{mandel20} to the optical and NIR light curves of 86 supernovae from CSP-I \citep{krisciunas17}. Using our hierarchical Bayesian framework to leverage the wide wavelength range covered by the CSP data ($B$--$H$ band; $\sim3500$--18500~\AA), we have been able to place strong constraints on the population distribution of host galaxy dust law $R_V$ values. Additionally, we have found that the Hubble residuals from distances estimated using the full optical and NIR light curves of the sample show evidence of a step with host galaxy stellar mass, and that this is insensitive to a variety of treatments of $R_V$ and its population distribution.

From the optical and NIR data of the low-to-moderate reddening SNe Ia (peak apparent $B-V\leq0.3$, consistent with the standard cut used in cosmological analyses), we infer an $R_V$ distribution with a population mean of $\mu_R=2.59\pm0.14$, and a population standard deviation of $\sigma_R=0.62\pm0.16$. Our estimate of the population mean is highly consistent with the estimate ($\mu_R=2.70\pm0.25$) obtained in \citet{thorp21}, who used the \textsc{BayeSN} model to study a completely independent cosmological SN Ia sample (Foundation DR1; \citealp{foley18,jones19}). Our estimate of the population standard deviation is not large, but is significantly different from zero, and indicates moderate scatter in SN Ia host galaxy $R_V$. This result is also consistent with the findings of \citet{thorp21}, who estimated that $\sigma_R\leq0.61$ with 95 per cent posterior probability. Here, the advantage of the near infrared data generally allows us to place stronger constraints on the small, but likely non-zero, size of the $R_V$ population distribution than \citet{thorp21}, who were only able to obtain upper limits.

Like \citet{thorp21}, we do not find strong evidence for a substantial mass dependence of the $R_V$ distribution. Restricting ourselves to the $B-V\leq0.3$ subset, we estimate $\mu_R=2.79\pm0.18$ for host galaxies less massive than the median ($10^{10.57}~\mathrm{M}_\odot$), and $\mu_R=2.35\pm0.27$ for those more massive. A similar result is obtained when splitting the sample at $10^{10}~\mathrm{M}_\odot$. As well as investigating correlations between $R_V$ and host mass, we also attempt to place constraints on the $R_V$ distribution among SNe Ia with redder apparent colours. Although previous works \citep{mandel11, burns14} have found evidence for an anti-correlation between $R_V$ and reddening or apparent colour, we generally do not find a strong effect of this nature \citep[c.f.\ also][]{rose22}. Whether splitting the sample into two ($B-V\leq0.3$ and $B-V>0.3$) or three ($B-V\leq0.3$, $0.3<B-V\leq0.7$, $B-V>0.7$) bins of apparent colour, we find consistency in the mean $R_V$ across the range of apparent colours. However, this inference is based on 11 SNe at $B-V>0.3$ (of which only 4 have $B-V>0.7$), so these results should be treated with caution -- a larger sample of highly reddened SNe Ia would be needed to properly characterise this effect. However, it is worth noting that this is the first study of optical--NIR colours that has been based on a continuous SED model -- the discrete-passband light curve models of \citet{mandel11} and \citet{burns14} relied on colour-dependent $K$ corrections to map the observed data  to the rest-frame. If these were inaccurate at high-reddening, it could explain why our results differ from \citet{mandel11} and \citet{burns14} in this regime.

We have found that our inferences of the $R_V$ distribution from the full optical and NIR light curves are almost unchanged when distance constraints based on redshifts and an assumed cosmology are imposed. This indicates that the rich colour information provided by the $B$ through $H$-band data is sufficient for constraining host galaxy dust properties, independently from the luminosity information provided by external distance estimates to the SNe, and thus independently from the cosmological model. An inspection of the \textsc{BayeSN} model colour curves (\S\ref{sec:why_onir}) reveals why this is the case -- when the full wavelength range is available, optical$-$optical colours provide information about reddening (i.e.\ $E(B-V)$), whilst optical$-$NIR colours are sensitive to extinction (i.e.\ $A_V$ or $A_B$). In combination, this allows for robust inference of $R_V$. When only optical$-$optical colours (e.g.\ $B-\{V,r,i\}$ are available, constraining $R_V$ with partial information is more challenging, and so the absolute magnitude information provided by independent, external estimates of distances must be invoked to break the degeneracy.

The sample we have analysed in this work is similar to that of \citet{burns14}, albeit using a new release of the CSP data, and with slightly different cuts. Although we adopt slightly different binning strategies, our inferences about the population mean $R_V$ are broadly consistent with the findings of \citet{burns14}. When assuming the \citet{fitzpatrick99} dust law and excluding fast decliners (\textsc{SNooPy} $s_{BV}<0.5$), they estimate $\mu_R=3.3\pm0.7$ and $\mu_R=2.6\pm0.4$ for SNe with $0.1<E(B-V)<0.3$ and $0.3<E(B-V)<0.4$, respectively. Alternatively, when fitting a Gaussian mixture to the full population of SNe with $B-V<0.5$, they estimate a dominant component with $\mu_R=2.3\pm0.3$. These values are all broadly consistent with our estimate of $\mu_R=2.59\pm0.14$ for our lowest-reddening bin (where $B-V\leq0.3$). They seem to find a stronger trend of mean $R_V$ with reddening \citep[see also][]{mandel11} than is seen in our own results, although given the uncertainties in the redder colour bins, the two sets of results are not inconsistent. That their Gaussian mixture model is dominated by a single component in all cases (see table 5 in \citealp{burns14}) agrees with our finding that the data do not show a strong preference for a non-Gaussian $R_V$ distribution (see Appendix \ref{app:altdistributions}). The most striking difference between our results and those of \citet{burns14} is in the inferred widths of our $R_V$ population distributions. For all colour bins, \citet{burns14} find a preference for wide distributions with $\sigma_R\gtrsim1$, although there is considerable uncertainty in some cases. Conversely our results point towards narrower distributions, with a high probability of $\sigma_R<1$ for all colour bins. The reason for this difference is not completely obvious, but could be related to prior choice (although see our Appendix \ref{app:sigmaRprior}), treatment of intrinsic colour, binning strategy, or differences in the sample selection and data.

A large number of CSP supernovae were also used in the more recent optical and NIR dust analysis of \citet{johansson21} (in combination with additional data from CfA/CfAIR, and their own iPTF sample). Taking the portion of their sample with $0.06<E(B-V)<0.5$, they estimate an $R_V$ distribution with mean $\mu_R=1.9$ and standard deviation $\sigma_R=0.9$. Splitting their sample by host mass at $10^{10}~\mathrm{M}_\odot$, they estimate $\mu_R=1.7$, $\sigma_R=0.8$ for high-mass hosts, and $\mu_R=2.2$, $\sigma_R=0.9$ for low-mass hosts. Their mean $R_V$ estimates are somewhat lower than our own in all cases, and their estimates of $\sigma_R$ tend to be wider. Their lower mean $R_V$ estimates may be due to their use of the \citet{cardelli89} dust law, which often seems to yield slightly lower estimates of $R_V$ (see, e.g.\ \citealp{burns14}, particularly table 1 therein). Alternatively, it could arise from the fact that unphysically negative values of $E(B-V)$ seem to be allowed in their fits (although they omit SNe with estimated $E(B-V)<0.06$~mag from their estimates of population mean). Allowing negative $E(B-V)$, whilst not allowing residual intrinsic colour scatter, could cause the $E(B-V)$ parameter to absorb a mixture of intrinsic colour and dust, thus pushing the estimated $R_V$ lower to capture an overall apparent colour--luminosity effect (similar to $\beta$ in the \citealp{tripp98}), rather than isolating a dust effect\footnote{This is also discussed in the context of \textsc{SNooPy} by \citet{jones22} -- see sections 3.1.1, 5.2.2, and appendix C therein. An extensive general discussion of the mixing of dust and intrinsic colour variation into an apparent colour--luminosity effect is provided by \citet{mandel17}.}. Their estimated low vs.\ high mass $\Delta\mu_R=0.5$ would not be inconsistent with our own results (e.g., for the $B-V<0.3$ subsample, we estimate $\Delta \mu_R=0.24\pm0.72$ when splitting at $10^{10}~\mathrm{M}_\odot$, or $\Delta \mu_R=0.43\pm0.32$ when splitting at the sample median host mass). Since \citet{johansson21} estimate $\sigma_R$ by taking the standard deviation of a collection of point estimates, their tendency towards larger estimates is unsurprising -- we have demonstrated in this work (\S\ref{sec:why_hb}) that such a strategy can lead to overestimation. The lack of accounting of residual intrinsic colour scatter when fitting the \textsc{SNooPy} \texttt{color\_model} \citep{burns11, burns14} is likely to be a contributing factor.

As well as investigating the dust law distribution in the CSP sample, we also estimate the Hubble residual mass steps obtained from distances estimated using optical-only, NIR-only, and optical+NIR data. Within this sample that is tilted towards more massive hosts (median mass $10^{10.57}~\mathrm{M}_\odot$), we find that the Hubble residuals obtained from our optical+NIR distances show a step of $0.056$--$0.064\pm0.026$~mag at the median host galaxy mass (when limited to the $B-V\leq0.3$ subset, and adopting a hierarchical treatment of $R_V$ and its population distribution). Our Hubble diagram total RMS is typically $\lesssim0.12$~mag. At $10^{10}~\mathrm{M}_\odot$ we estimate a similar sized, but more uncertain, mass step, whilst at the maximum likelihood step location ($\sim10^{10.8}~\mathrm{M}_\odot$) we estimate a larger step $0.082$--$0.090\pm0.028$~mag. The step size is largely insensitive to the treatment of host galaxy $R_V$.  We infer distances under fitting configurations where $R_V$ is fixed to a fiducial value for all SNe ($R_V=2.89$; \citealp{mandel20}), allowed to vary from SN to SN within an inferred population distribution (under various forms of this distribution), or is inferred independently for each SN under one of two weakly informative priors. Using Hubble residuals inferred from optical data only, we find similar or larger mass steps at all locations, and a larger Hubble diagram scatter that is more sensitive to the treatment of $R_V$. When the Hubble residuals are based on NIR-only estimates of distance, we estimate a step of $0.050\pm0.032$~mag at the median host mass, and $0.084\pm0.035$~mag at the MLE step location. For the full sample, the NIR-only Hubble diagram total RMS is 0.137~mag, smaller than for any of the optical-only results. For the subsample with NIR data at maximum, we obtain an even smaller Hubble diagram RMS of 0.091~mag. Like the results of \citet{ponder20} and \citet{jones22}, we estimate a positive, but fairly uncertain, NIR mass step. Our step sizes' credible intervals would be consistent with both the larger steps found by \citet{uddin20} in $YJH$ and \citet{johansson21} in $Y$, or the smaller size estimated by \citet{johansson21} in $JH$. Larger samples of SN Ia in the NIR will provide more stringent constraints.

In summary, we have been able to leverage the high quality optical and NIR data from CSP-I to place strong constraints on the distribution of host galaxy dust law $R_V$ values in SN Ia hosts. With this independent sample, we have upheld the conclusions of \citet{thorp21} -- namely that the lines of sight to low reddening ($B-V\leq0.3$) SNe Ia typical of a cosmological sample have an $R_V$ distribution of small to moderate width, with limited dependence on host galaxy stellar mass. Contrary to other recent work \citep{johansson21}, we have found that a mass step persists in the Hubble residuals of SNe Ia distances estimated from the full optical and NIR light curves, and that this is robust to a variety of Bayesian treatments of host galaxy $R_V$ and its population distribution. When estimating distances from the NIR light curves alone, our results depend on the step location. We find a strong NIR-only step at the maximum likelihood step location, comparable to the equivalent steps in the optical+NIR or optical-only Hubble residuals. Our estimated NIR Hubble residual step at the median host galaxy mass is more tenuous. A weaker NIR mass step could be indicative of a wavelength-dependent correlation between intrinsic SN Ia luminosity and host mass, or an SN--host correlation that is driven by a combination of dust and intrinsic factors. Further work is needed to properly understand this. However, irrespective of the size of any NIR mass step, our finding that the optical+NIR Hubble residuals consistently show a step even when marginalising over the dust law distribution suggests that dust is unlikely to be the only driver of observed SN--host correlations.

We have affirmed here the value of combining optical and NIR data for investigating the dust in SN Ia host galaxies and understanding SN--host correlations. Of critical importance to future investigation is the growth of an SN Ia sample with NIR light curves, and a more representative host galaxy population than was observed by the CSP-I \citep{krisciunas17} and CfAIR2 \citep{woodvasey08,friedman15} surveys. Such data are beginning to become available from the RATIR followup of the iPTF survey \citep{johansson21}, with more on expected on the horizon from CSP-II \citep{phillips19}. The Dark Energy, $\mathrm{H}_0$, and peculiar Velocities using Infrared Light from Supernovae \citep[DEHVILS;][]{konchady22} survey on the UK Infrared Telescope (UKIRT), and $y$-band Pan-STARRS data from the Young Supernova Experiment \citep[YSE;][]{jones21} will further enable future investigations. Expanding the sample of high redshift SNe Ia with rest-frame NIR observations \citep[e.g.][]{jones22} will also be of great importance for constraining any potential redshift-evolution of SN Ia host galaxy dust properties.

The ongoing development of robust statistical techniques for modelling dust and SN--host correlations, combined with the continual expansion of the NIR dataset, and the investigation of novel approaches for constraining SN Ia dust and intrinsic properties (e.g.\ the use of SN Ia siblings; \citealp{biswas22}; Ward et al.\ in prep.), should enable us to finally address one of the outstanding questions of SN Ia cosmology. This will be vital to ensuring that our understanding of SN Ia dust, and correlations between SNe Ia and their hosts, do not limit our ability to capitalise on the data from the Vera C.\ Rubin Observatory and \textit{Nancy Grace Roman Space Telescope}.

\section*{Acknowledgements}
We thank the anonymous referee for their encouraging review and helpful comments. We thank Gautham Narayan for his valuable contributions to the development and implementation of the \textsc{BayeSN} model, and for supporting our access to the Illinois Campus Cluster. We thank Sam M.\ Ward and Suhail Dhawan for helpful discussions. This work made use of files and metadata compiled by David O.\ Jones, Andrew S.\ Friedman, and Arturo Avelino from public datasets. We thank the Carnegie Supernova Project team for making their SN Ia photometry and host galaxy mass estimates publicly available.

ST was supported by the Cambridge Centre for Doctoral Training in Data-Intensive Science funded by the UK Science and Technology Facilities Council (STFC). KSM acknowledges funding from the European Research Council under the European Union’s Horizon 2020 research and innovation programme (ERC Grant Agreement No. 101002652). This project has been made possible through the ASTROSTAT-II collaboration, enabled by the Horizon 2020, EU Grant Agreement No. 873089.  

This work made use of the Illinois Campus Cluster, a computing resource that is operated by the Illinois Campus Cluster Program (ICCP) in conjunction with the National Center for Supercomputing Applications (NCSA) and which is supported by funds from the University of Illinois at Urbana-Champaign.

\section*{Data Availability}
 

All photometric data used in this paper are publicly available from CSP Data Release 3 \citep{krisciunas17}. Host galaxy stellar masses are publicly available from \citet[table C1]{uddin20}.



\bibliographystyle{mnras}
\bibliography{main} 




\appendix
\section{Hyperprior Sensitivity of Dust Law Population Distribution}
\label{app:sigmaRprior}
As in \citet[appendix B]{thorp21}, we consider here the sensitivity of our $R_V$ population distribution inferences to our choice of prior. In particular, we focus on the choice of prior on the standard deviation parameter(s), $\sigma_R$. We rerun our partial pooling analyses, where $R_V$ is either unbinned or binned by $B-V$ colour, using several alternative choices of prior on $\sigma_R$. As well as our default prior, $\sigma_R\sim\text{Half-}N(0,2^2)$, we follow \citet{thorp21} in testing a tighter half-normal prior, $\sigma_R\sim\text{Half-}N(0,1^2)$, and a uniform prior, $\sigma_R\sim U(0,4)$. The former promotes a higher degree of shrinkage than our default choice, whilst the latter specifies equal preference for any $\sigma_R<4$ (i.e. the range over which we place 95 per cent prior probability in our default choice). In addition to these, we expand on \citet{thorp21} to consider an improper flat prior, $\sigma_R\sim U(0,\infty)$, which should yield a proper posterior (see \citealp{gelman06}; \citealp{gelman14}, \S5.7) even for the smallest colour bin (where $N=4$). We also test the hierarchical prior from \citet[\citetalias{burns14}]{burns14}\footnote{Their appendix A.1.2 suggests that they adopt this prior for both their Gaussian mixture $R_V$ population distribution model, and for their $E(B-V)$ binned model inspired by \citet{mandel11}. However, it is not totally explicit that they adopt this prior in the latter case.},
\begin{align}
    w^2 &\sim U(0,\infty),\label{eq:b14wprior}\\
    \sigma_R^2 | w^2 &\sim \text{Inv-}\chi^2(1, w^2)\label{eq:b14sigmaprior},
\end{align}
where $\sigma_R^2$ has a scaled inverse-$\chi^2$ prior with one degree of freedom, and a scale parameter, $w$, whose square has an improper uniform prior. Since the inverse-$\chi^2$ distribution is zero-avoiding (i.e. places zero probability density at $\sigma_R = 0$ and therefore a priori excludes it and penalises small values of $\sigma_R$), and the improper prior on $w^2$ will concentrate at $w^2\to\infty$, we expect this prior choice will encourage high estimates of $\sigma_R$. It will certainly prevent the estimates of $\sigma_R$ from shrinking towards zero, even if the number of groups being considered (five colour bins in \citealp{burns14}) is enough to constrain $w^2$. This prior choice may explain the results of \citet{burns14}, who estimated wide $R_V$ population distributions (with posterior mean $\sigma_R\gtrsim1$) in all of their reddening bins (see their table 6).

\begin{table*}
    \centering
    \begin{threeparttable}
        \caption{Sensitivity analysis of the constraints on the dust $R_V$ population std.\ dev. hyperparameter $\sigma_R$, from full optical and NIR light curves under alternative hyperprior choices.}
        \label{tab:sigma_RV_priors}
        \begin{tabular}{l c c c c c c}\toprule
            & & \multicolumn{5}{c}{$\sigma_R$\tnote{a}}\\
            \cmidrule(l){3-7}
            Binning\tnote{b} & Subsample & $\text{Half-}N(0,2^2)$\tnote{c} & $\text{Half-}N(0,1^2)$ & $U(0,4)$ & $U(0,\infty)$ & \citetalias{burns14}\tnote{d}\\\midrule
            Unbinned & - & $0.34\pm0.11$ & $0.33\pm0.11$ & $0.33\pm0.11$ & $0.34\pm0.11$ & $0.38\pm0.10$\\
            2-bin & $B-V\leq0.3$ & $0.62\pm0.16$ & $0.54\pm0.16$ & $0.55\pm0.16$ & $0.59\pm0.16$ & $0.51\pm0.15$\\
            & $B-V>0.3$ & $0.17\pm0.10$ & $0.17\pm0.10$ & $0.17\pm0.10$ & $0.18\pm0.11$ & $0.24\pm0.12$\\
            3-bin & $B-V\leq0.3$ & $0.57\pm0.20$ & $0.57\pm0.15$ & $0.58\pm0.17$ & $0.57\pm0.16$ & $0.52\pm0.16$\\
            & $0.3<B-V\leq0.7$ & 0.34 (1.02) & 0.29 (0.56) & 0.32 (0.68) & 0.33 (0.72) & $0.30\pm0.18$\\
            & $B-V>0.7$ & 0.31 (0.76) & 0.28 (0.62) & 0.28 (0.65) & 0.31 (0.83) & $0.30\pm0.21$\\
            \bottomrule
        \end{tabular}
        \begin{tablenotes}
            \item [a] Different columns correspond to different $\sigma_R$ hyperprior choices. Values quoted are either posterior mean $\pm$ standard deviation, or 68 (95)\% upper bounds.
            \item [b] Colour binning strategy.
            \item [c] Default prior choice (reproduced from Table \ref{tab:pop_RV_params_optnir}).
            \item [d] \citet[\citetalias{burns14}]{burns14} hierarchical prior (Eq. \ref{eq:b14wprior}, \ref{eq:b14sigmaprior}).
        \end{tablenotes}
    \end{threeparttable}
\end{table*}

Table \ref{tab:sigma_RV_priors} lists our posterior inferences on $\sigma_R$ under the alternative prior choices. For the 1- or 2-bin models, the $\text{Half-}N(0,1^2)$, $U(0,4)$, and $U(0,\infty)$ priors have negligible impact on our $\sigma_R$ posteriors compared to our default $\text{Half-}N(0,2^2)$ choice. For the 3-bin model, there is some sensitivity in the tail of the posterior, but it is fairly small. For example, in switching from the $\sigma_R\sim \text{Half-}N(0,2^2)$ to the $\sigma_R\sim \text{Half-}N(0,1^2)$ prior, the 95th percentile of the posterior for the $0.3<B-V\leq0.7$ colour bin shifts inwards from $\sigma_R=1.02$ to $\sigma_R=0.56$. For the $B-V>0.7$ bin, the 95th percentile shifts from $\sigma_R=0.76$ to $\sigma_R=0.62$ under the same change of prior. The shift in the 68th percentile is smaller, only $\lesssim0.05$ for both colour bins. Under the \citetalias{burns14} prior (Eq.\ \ref{eq:b14wprior}, \ref{eq:b14sigmaprior}), the posterior estimates of $\sigma_R$ are consistent with those from the other prior choices. However, the posteriors peak away from zero for all bins, something that isn't true for the other priors (which do not a priori exclude zero). The prior scale hyperparameter, $w$, has a very similar posterior for the model configurations with two or three colour bins. For the 3-bin configuration, we estimate that $0.09\leq w\leq0.82$ with 95 per cent posterior probability. For the 1-bin configuration, the prior scale is less constrained, with the 95 per cent credible interval covering $0.16\leq w\leq1.34$.

\begin{figure}
    \centering
    \includegraphics[width=\linewidth]{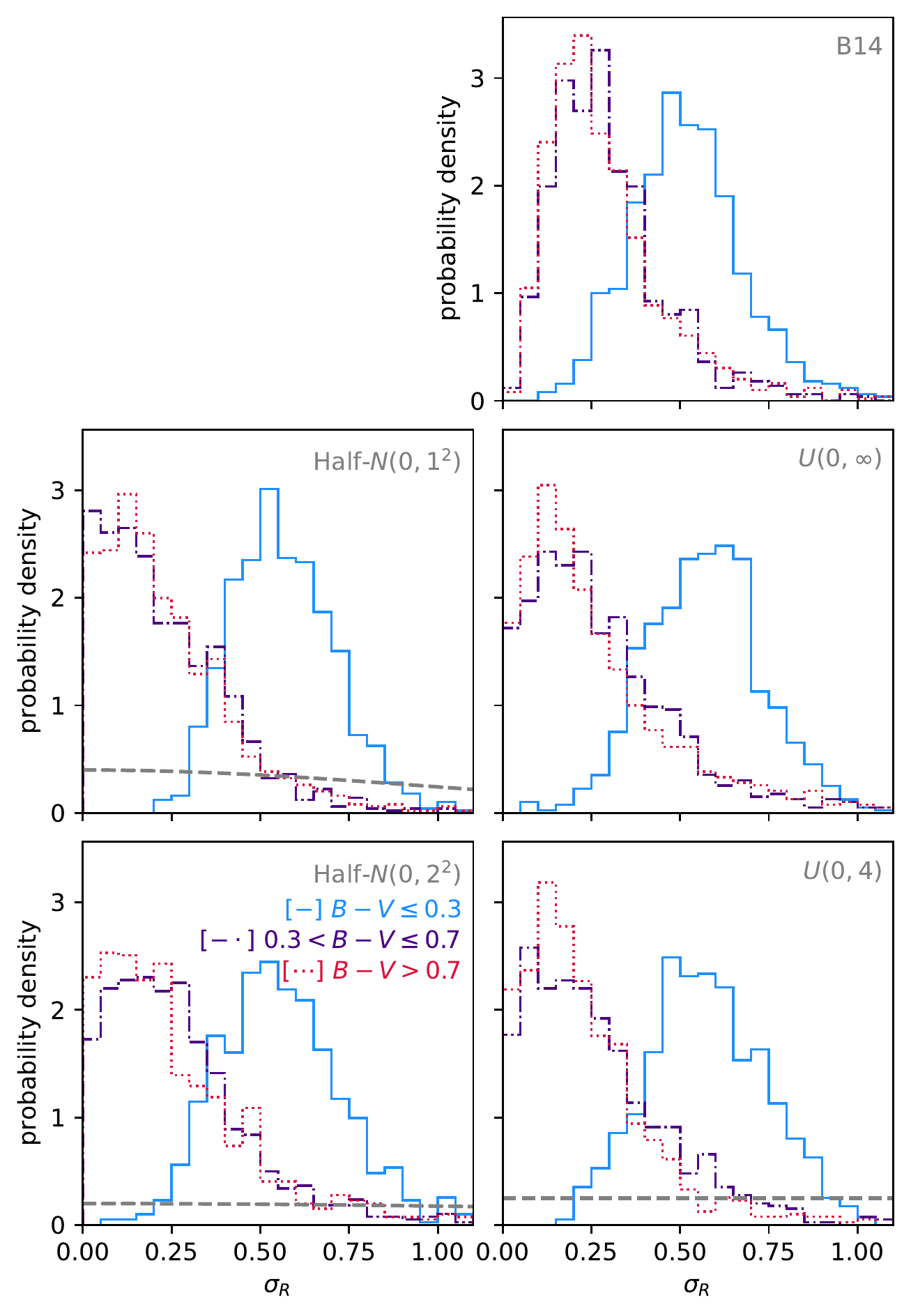}
    \caption{Histograms of $\sigma_R$ posterior samples under different choices of prior for the model configuration where an $R_V$ population distribution is inferred within each of 3 colour bins. Axes are the same for all panels. Coloured lines depict the $\sigma_R$ posteriors in different colour bins. Grey lines show the priors. For the \citetalias{burns14} prior, the prior scale parameter, $w^2$, is a free parameter, so no prior is plotted. The histogram bins have a width of 0.05.}
    \label{fig:sigmaRprior}
\end{figure}

Figure \ref{fig:sigmaRprior} shows the posterior distribution of $\sigma_R$ in different colour bins (under the 3-bin model) for different choices of prior. The posteriors are depicted via histograms of the MCMC samples, rather than kernel density estimates, and with a more zoomed in $\sigma_R$ axis than Figure \ref{fig:distributions}. This is to highlight the behaviour at low $\sigma_R$. The lower left panel uses our fiducial prior, $\sigma_R\sim\text{Half-}N(0,2^2)$, and is equivalent to the lower left panel of Figure \ref{fig:distributions}. The zero-avoidance induced by the \citetalias{burns14} prior is clearly visible in the upper right panel -- none of the other prior choices yield posteriors that exhibit this behaviour.

Overall, the only slight differences resulting from the prior choices we have tried here indicate that our posterior results are relatively insensitive and robust to them.
 
\section{Alternative Forms for the Dust Law Population Distribution}
\label{app:altdistributions}
Here, we present results from photometric distance fits with partial pooling of $R_V$ (Section \ref{sec:fittingconfigs}~\ref{itm:pp}), but with alternative forms for the $R_V$ population distribution than our fiducial choice of a truncated normal (Eq.\ \ref{eq:rvprior}). Specifically, we repeat our analysis assuming either a skew normal\footnote{We define $X\sim\text{Skew-}N(\xi, \omega, \alpha)$ as a random variable drawn from a skew normal distribution, with location parameter $\xi$, scale parameter $\omega$, and skewness parameter $\alpha$. The probability density function is\begin{equation*}P(X|\xi,\omega,\alpha) = \frac{2}{\omega}\phi\left(\frac{X-\xi}{\omega}\right)\Phi\left(\alpha\frac{x-\xi}{\omega}\right),\end{equation*} where $\phi(z)$ and $\Phi(z)$ are respectively the PDF and CDF of a standard normal random variable $z$.} population distribution,
\begin{equation}
    R_V^s \sim \text{Skew-}N(\xi_R, \omega_R, \alpha_R) \text{ for } R_V^s \geq 0.5,
    \label{eq:skewn}
\end{equation}
or a Student's $t$\footnote{We define $X\sim\text{Student-}t(\nu, \mu, \sigma)$ as a random variable drawn from a Student's $t$-distribution, with location parameter $\mu$, scale parameter $\sigma$, and degrees-of-freedom $\nu$. The probability density function is\begin{equation*}P(X|\nu,\mu,\sigma) = \frac{\Gamma[(\nu+1)/2]}{\Gamma(\nu/2)\sqrt{\nu\pi}\sigma}\left[1 + \frac{1}{\nu}\left(\frac{X-\mu}{\sigma}\right)^2\right]^{-(\nu+1)/2},\end{equation*} where $\Gamma(z)$ is the gamma function.} population distribution,
\begin{equation}
    R_V^s \sim \text{Student-}t(\nu_R, \mu_R, \sigma_R) \text{ for } R_V^s \geq 0.5,
    \label{eq:studentt}
\end{equation}
both truncated to $R_V\geq0.5$\footnote{Let a random variable, $X$, follow an untruncated probability distribution with PDF $f(X)$ and CDF $F(X)$. Let a second random variable, $Y$, follows the same probability distribution, but truncated on its lower tail so $Y\geq a$. The probability density function for $Y$ will be \begin{equation*}P(Y) = \begin{cases}f(Y)/[1 - F(a)] &\text{ for } Y \geq a\\0 &\text{ elsewhere}\end{cases}.\end{equation*}}. These distributions were chosen to probe specific kinds of non-Gaussianity (skewness, or heavy-tailed-ness, respectively). Both can revert to a normal distribution as a special case, with $\text{Skew-}N(\xi, \omega, \alpha) \to N(\xi, \omega^2)$ for $\alpha=0$, and $\text{Student-}t(\nu,\mu,\sigma)\to N(\mu,\sigma^2)$ as $\nu\to\infty$.

When adopting these alternative $R_V$ distributions, we use the same hyperpriors \citep[from][]{thorp21} on their location and scale parameters as we do when assuming a truncated normal $R_V$ distribution, i.e. $\text{location} \sim U(1,5)$, and $\text{scale} \sim \text{Half-}N(0,2^2)$. For the skew normal distribution, we adopt a Gaussian prior on the skewness parameter, $\alpha_R\sim N(0,2^2)$, since this allows a fairly substantial skew without preference for either direction. For the $t$-distribution's degrees-of-freedom parameter, we follow \citet{juarez10} in adopting a Gamma distribution prior\footnote{This choice is weakly informative, and covers a wide range of possibilities, with the 1st and 99th percentiles falling at $\nu_R=1.5$ (where the $t$-distribution will be nearly a Cauchy) and $\nu_R=66.4$ (where the $t$-distribution will be practically Gaussian), respectively. We expect this simple choice will be sufficient for our purposes. For alternatives, see e.g.\ \citet{fonseca08, juarez10, fruhwirthschnatter10, rubio15, simpson17}.}, $\nu_R\sim\gammadist(\alpha=2.0,\beta=0.1)$.

\begin{figure}
    \centering
    \includegraphics[width=\linewidth]{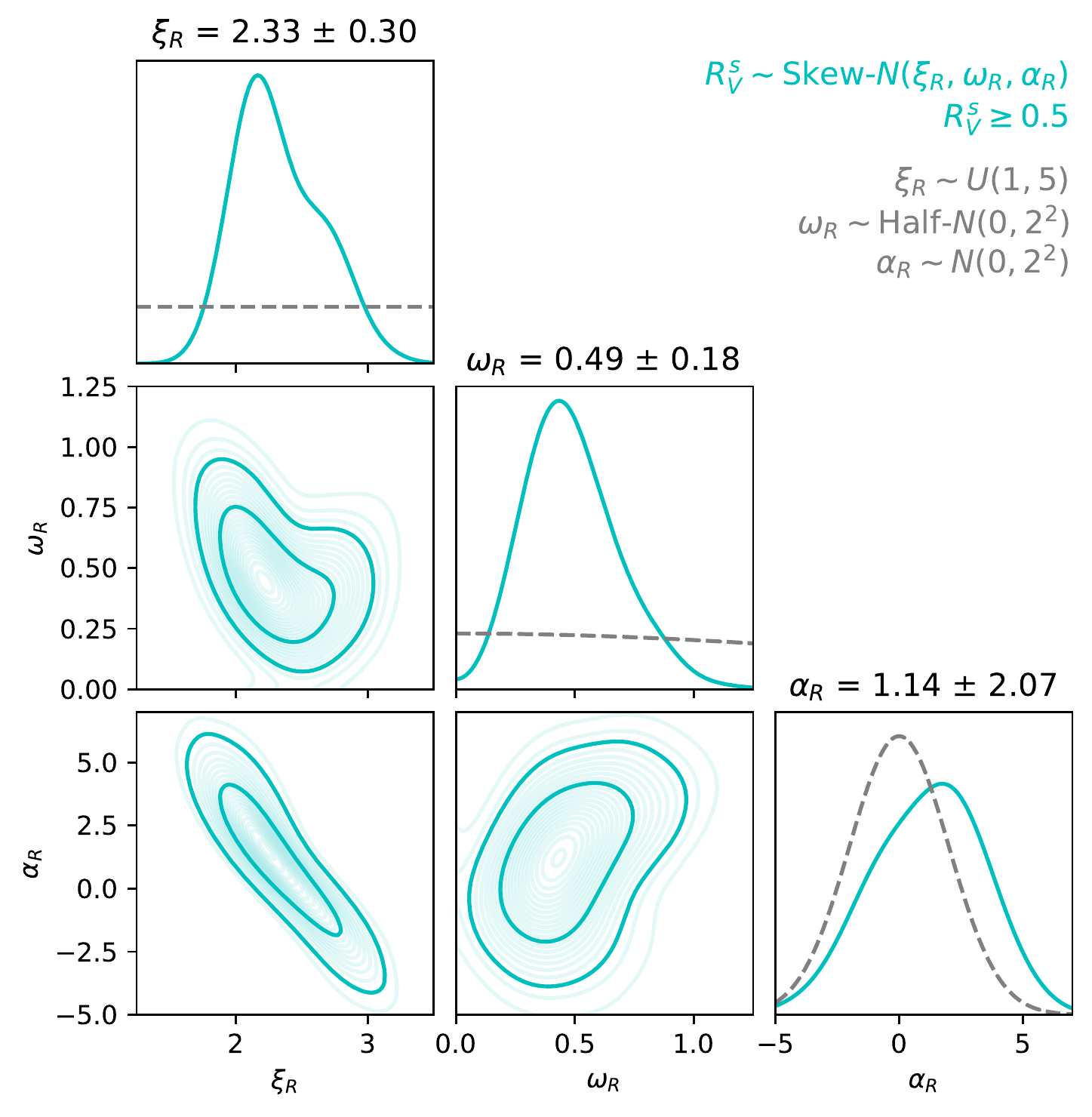}
    \includegraphics[width=\linewidth]{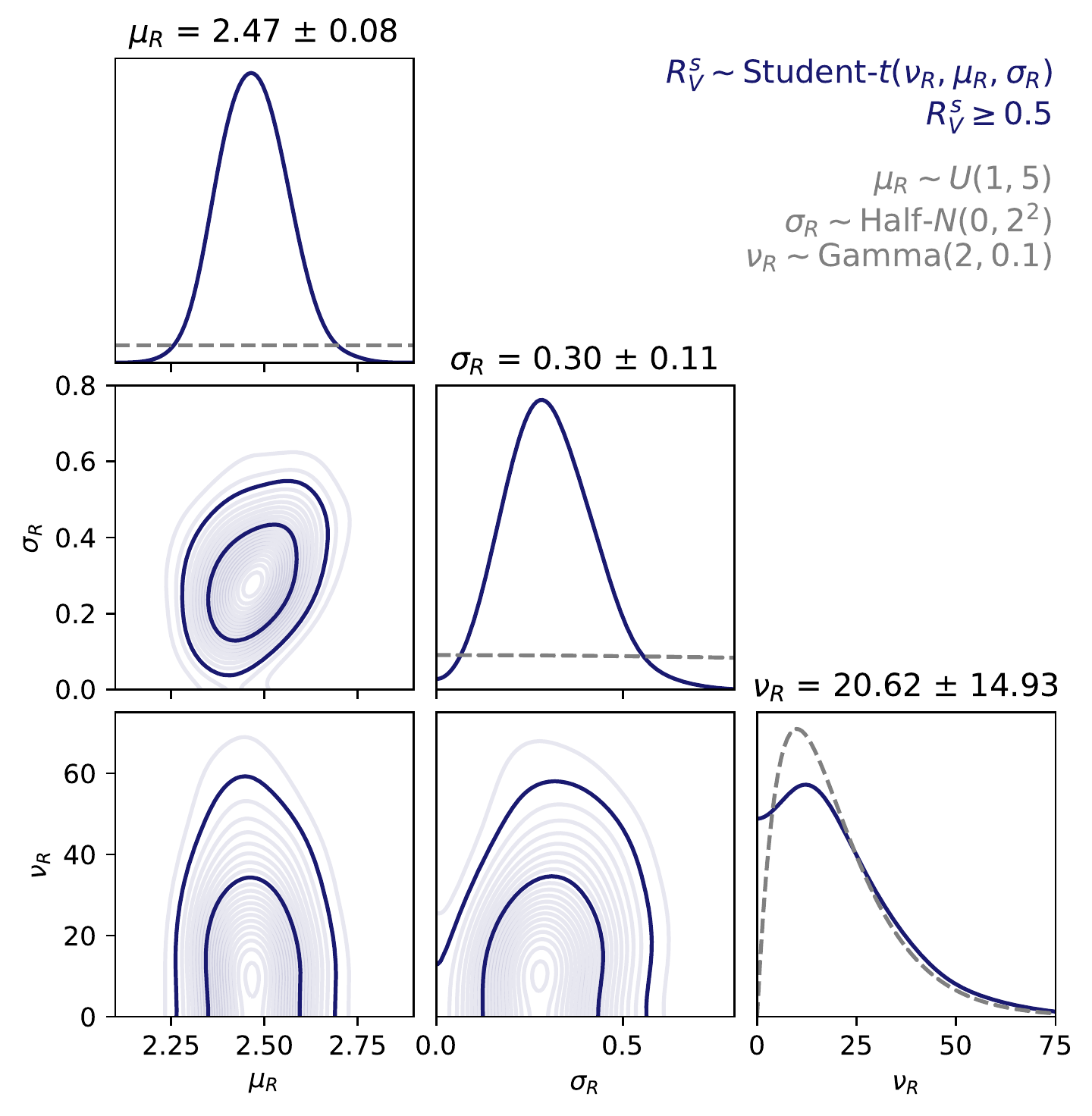}
    \caption{Posterior inferences of dust $R_V$ population distribution parameters for the two alternative distributional forms. Coloured lines depict the posteriors. Faint grey lines on the 1D marginals show the priors. (top panel) Skew normal $R_V$ population distribution (Eq.\ \ref{eq:skewn}). (bottom panel) Student's $t$ population distribution (Eq.\ \ref{eq:studentt}).}
    \label{fig:altdistributions}
\end{figure}

Under these two models, our constraints on any non-Gaussianity of the $R_V$ distribution are limited. For the skew normal population distribution, we estimate a location parameter $\xi_R=2.33\pm0.30$, and a scale parameter $\omega_R=0.49\pm0.18$. The skewness parameter shows a weak preference for a positively skewed distribution, with $\alpha_R=1.14\pm2.08$, although both directions of skew are permitted in the posterior with high probability. The upper panel of Figure \ref{fig:altdistributions} shows our joint posterior over the three parameters of the skew normal distribution. Although the posterior mean of the location parameter, $\xi_R$, is lower than the posterior estimate of $\mu_R$ for the Gaussian population distribution (c.f. Table \ref{tab:pop_RV_params_optnir}, and the upper left panels of Fig.\ \ref{fig:distributions}), it is worth noting that $\xi_R$ does not equal the implied distribution mean unless $\alpha_R=0$. In fact, for each MCMC sample, $i$, we can estimate the expected value of $R_V$, $\mathbb{E}(R_V|\xi_R^i,\omega_R^i,\alpha_R^i)$, implied by the population distribution parameters $(\xi_R^i, \omega_R^i, \alpha_R^i)$ under the skew normal population distribution. Doing this, we estimate that $\mathbb{E}(R_V|\xi_R,\omega_R,\alpha_R)=2.50\pm0.09$. This is almost perfectly consistent with our posterior estimate of $\mu_R=2.48\pm0.08$ for the mean of the Gaussian population distribution.

For the Student's $t$-distribution, we estimate a location parameter of $\mu_R=2.47\pm0.08$, and a scale of $\sigma_R=0.30\pm0.11$. These values are quite consistent with the population mean and standard deviation parameters we estimate when assuming a Gaussian population distribution (see Table \ref{tab:pop_RV_params_optnir} and Fig.\ \ref{fig:distributions}). The degrees-of-freedom parameter, $\nu_R$ has a posterior very close to the prior, suggesting that there is little preference for any particular level of heavy-tailed-ness in this sample. The lower panel of Figure \ref{fig:altdistributions} shows the posterior distribution of the parameters of the Student's $t$ $R_V$ population distribution.

We find that marginalising over these alternative $R_V$ population distributions when inferring distances has very little impact on the estimated mass step in the resulting Hubble residuals.  Across the range of possible mass step locations, our estimated Hubble residual steps are highly consistent with those estimated when our fiducial choice of a Gaussian $R_V$ population distribution was adopted.

\section{Distribution of Dust Laws: Using Optical Data Only}
\label{sec:optdustlaws}

\begin{table*}
    \centering
    \begin{threeparttable}
        \caption{Same as Table \ref{tab:pop_RV_params_optnir}, but dust $R_V$ population constraints using optical-only ($BVri$) data with external distance information applied.}
        \label{tab:pop_RV_params_opt}
        \begin{tabular}{l c c c c}\toprule
            Binning & Subsample & $N$ & $\mu_R$ & $\sigma_R$\\\midrule
            Unbinned & - & 86 & $2.53\pm0.15$ & $0.51\pm0.14$\\
            Colour (2-bin) & $B-V\leq0.3$ & 75 & $2.72\pm0.17$ & $0.40\pm0.21$\\
            & $B-V>0.3$ & 11 & $2.14\pm0.30$ & $0.60\pm0.34$\\
            Colour (3-bin) & $B-V\leq0.3$ & 75 & $2.73\pm0.15$ & $0.35\pm0.19$\\
            & $0.3<B-V\leq0.7$ & 7 & $2.39\pm0.36$ & $0.74\pm0.50$\\
            & $B-V>0.7$ & 4 & $1.83\pm0.29$ & $0.55\pm0.47$\\
            Mass ($10^{10}~M_\odot$) & low & 15 & $2.79\pm0.49$ & $1.29\pm0.62$\\
            & high & 71 & $2.47\pm0.15$ & $0.45\pm0.14$\\
            Mass ($10^{10.57}~M_\odot$) & low & 43 & $2.76\pm0.19$ & $0.60\pm0.18$\\
            & high & 43 & $2.26\pm0.23$ & $0.52\pm0.35$\\
            \bottomrule
        \end{tabular}
    \end{threeparttable}
\end{table*}

The most reliable inference of $R_V$ and its population distribution is achieved when leveraging the full optical and NIR ($BVriYJH$) light curves of our sample. Nevertheless, we can also attempt our analysis whilst using only the optical ($BVri$) light curves. This Appendix presents these results, with the use of external distance information. Without the use of external distance constraints, population $R_V$ inferences are not very reliable, for the reasons discussed in Section \ref{sec:why_onir}. Under these conditions, $R_V$ is constrained only by colour information (i.e. relative differences in magnitude between passbands). Optical colours by themselves do not effectively constrain individual $R_V$ values nor their population distribution for most of the sample (c.f.\ Fig.~\ref{fig:colour_curves}). When external distance constraints are imposed, $R_V$ is informed by colour--magnitude information, rather than colours alone. Stronger constraints are possible from the optical light curves under these circumstances.

Table \ref{tab:pop_RV_params_opt} lists our posterior estimates of the $R_V$ population distribution parameters $(\mu_R,\sigma_R)$ from fitting optical ($BVri$) data with external distance constraints imposed. For all configurations considered, our estimates of the population mean, $\mu_R$, are consistent to within $\lesssim2\sigma$ with the equivalent estimates made using the full $BVriYJH$ data. The same is true for our estimates of the population standard deviation parameter, $\sigma_R$.

The colour-binned results from fitting only the optical data would seem to suggest a stronger trend of mean $R_V$ with colour/reddening \citep[\`a la][]{mandel11,burns14} than the results from the optical + NIR analysis. This is particularly apparent in the 3-bin configuration, where the population mean $R_V$ estimates for the lowest- ($B-V\leq0.3$) and highest-reddening ($B-V>0.7$) supernovae are $\sim2.8\sigma$ apart when only the $BVri$ data are considered. Although this appears to be drastically different to the conclusion from the analysis of the $BVriYJH$ data, where the same colour bins differ by only $\sim1\sigma$ in estimated mean $R_V$, the actual parameter estimates from the two analyses differ by at most $\sim1.4\sigma$ (with the biggest difference being between the $\mu_R$ estimates in the very sparsely populated $B-V>0.7$ bin).


\bsp	
\label{lastpage}
\end{document}